\documentclass{emulateapj}
\usepackage{graphicx}
\usepackage{amsmath}
\def\kms{{\ }{\rm km}\,{\rm s}^{-1}}
\def\LCDM{$\Lambda$CDM}

\begin{document}

\submitted{Accepted by ApJL on January 26, 2013}
           
\slugcomment{{\it Accepted by ApJL on January 26, 2013}}
                  
\shortauthors{KAZANTZIDIS, {\L}OKAS, \& MAYER}  

\shorttitle{Tidal Stirring of Disky Dwarfs with Shallow Dark Matter
  Density Profiles}

\title{Tidal Stirring of Disky Dwarfs with Shallow Dark Matter Density
  Profiles: Enhanced Transformation into Dwarf Spheroidals}

\author{Stelios Kazantzidis\altaffilmark{1}, 
  Ewa L. {\L}okas\altaffilmark{2}, and
  Lucio Mayer\altaffilmark{3}}

\altaffiltext{1}{Center for Cosmology and Astro-Particle Physics; and
  Department of Physics; and Department of Astronomy, The Ohio State
  University, Columbus, OH 43210, USA; stelios@mps.ohio-state.edu}
\altaffiltext{2}{Nicolaus Copernicus Astronomical Center, 00-716
  Warsaw, Poland} 
\altaffiltext{3}{Institute for Theoretical Physics, University of Z\"urich, 
  CH-8057 Z\"urich, Switzerland} 

\begin{abstract}

  The origin of dwarf spheroidal galaxies (dSphs) in the Local Group
  (LG) remains an enigma. The tidal stirring model posits that
  late-type, rotationally-supported dwarfs resembling present-day
  dwarf irregular (dIrr) galaxies can transform into dSphs via
  interactions with Milky Way-sized hosts. Using collisionless
  $N$-body simulations, we investigate for the first time how tidal
  stirring depends on the dark matter (DM) density distribution in the
  central stellar region of the progenitor disky dwarf. Specifically,
  we explore various asymptotic inner slopes $\gamma$ of the dwarf DM
  density profiles ($\rho \propto r^{-\gamma}$ as $r \rightarrow 0$).
  For a given orbit inside the primary galaxy, rotationally-supported
  dwarfs embedded in DM halos with core-like density distributions
  ($\gamma = 0.2$) and mild density cusps ($\gamma = 0.6$) demonstrate
  a substantially enhanced likelihood and efficiency of transformation
  into dSphs compared to their counterparts with steeper DM density
  profiles ($\gamma = 1$).  Such shallow DM distributions are akin to
  those of observed dIrrs, highlighting tidal stirring as a plausible
  model for the origin of the morphology-density relation in the LG.
  When $\gamma <1$, a single pericentric passage can induce dSph
  formation and disky dwarfs on low-eccentricity or large-pericenter
  orbits are able to transform into dSphs; these new results allow the
  tidal stirring model to explain the existence of virtually all known
  dSphs across a wide range of distances from their hosts. A subset of
  rotationally-supported dwarfs initially embedded in DM halos with
  shallow density profiles are eventually disrupted by the primary
  galaxy; those that survive as dSphs are generally on orbits that are
  biased towards lower eccentricities and/or larger pericenters
  relative to those of typical cold dark matter (CDM) satellites. The
  latter could explain the rather peculiar orbits of several classic
  LG dSphs such as Fornax, Leo I, Tucana, and Cetus. We conclude that
  tidal stirring constitutes a prevalent evolutionary mechanism for
  shaping the nature of dwarf galaxies within the currently favored
  CDM cosmological paradigm.

\end{abstract}


\section{Introduction}
\label{sec:introduction}

The dwarf spheroidal (dSph) satellites in the Local Group (LG) are the
faintest and most dark-matter (DM) dominated galaxies known
\citep[e.g.,][]{Mateo98,Tolstoy_etal09}. These intriguing objects
exhibit no appreciable gas components and tend to be clustered around
the massive spiral galaxies Milky Way (MW) and M31. Furthermore, dSphs
are characterized by pressure-supported, spheroidal stellar components
\citep[e.g.,][]{Mateo98} and a wide diversity of star formation
histories \citep[e.g.,][]{Orban_etal08}. Despite recent advances in
our understanding of dSphs, a conclusive model for their origin has
yet to emerge \citep[e.g.,][]{Kravtsov10}.

One class of models attempts to explain the formation of dSphs via
environmental processes, including tidal and ram pressure stripping
\citep[e.g.,][]{Einasto_etal74,Faber_Lin83,Mayer_etal01,Kravtsov_etal04,
  Mayer_etal06,Mayer_etal07,Klimentowski_etal09,Kazantzidis_etal11a,Lokas_etal11},
mergers between dwarf galaxies
\citep{Kazantzidis_etal11b,Yozin_Bekki12}, and resonant stripping
\citep{D'Onghia_etal09}. In this context, the ``tidal stirring'' model
of \citet{Mayer_etal01} postulates that interactions with MW-sized
host galaxies can transform late-type, rotationally-supported dwarfs
resembling present-day dwarf irregular (dIrr) galaxies into
pressure-supported, spheroidal stellar systems with the kinematic and
structural properties of the {\it classic} LG dSphs (see
\citealt{Lokas_etal12a} for an extension of this model to the
newly-discovered ultra-faint dSphs).

Previous work on tidal stirring has almost exclusively adopted cuspy,
\citet[][hereafter NFW]{Navarro_etal96} density profiles to model the
DM halos of the progenitor disky dwarfs. Yet, rotation-curve modeling
of observed dIrrs \citep[e.g.,][]{Weldrake_etal03,Oh_etal11}, which
constitute the most plausible predecessors of dSphs according to the
tidal stirring model, favors DM halos with significantly shallower
inner density slopes.  Such shallow DM density distributions are also
supported by hydrodynamical cosmological simulations of dwarf galaxy
formation \citep{Governato_etal10} and recent analytic models for the
origin of core-like DM density profiles in dwarfs
\citep{Pontzen_Governato12}.

Given that tidal stirring involves a combination of tidally-induced
dynamical instabilities (e.g., bars) and impulsive tidal heating of
the stellar component, the cuspiness of the dwarf DM density
distribution in the central region where the stars reside should play
a crucial role in this complex transformation mechanism.  Indeed,
disky dwarfs with shallow DM density profiles are characterized by
lower central densities and, correspondingly, longer internal
dynamical times compared to their counterparts with steeper DM density
distributions.  Systems with longer dynamical times respond more
impulsively to external tidal perturbations and suffer stronger tidal
heating \citep[e.g.,][]{Gnedin_Ostriker99}.  Rotationally-supported
dwarf galaxies with progressively shallower DM density profiles may
thus experience increasingly probable and efficient transformations
into dSphs.  Nonetheless, there has been no {\it quantitative} work
aimed at investigating this qualitative expectation.

Here we explore this issue via a series of tidal stirring simulations
of disky dwarfs embedded in DM halos with different inner density
distributions. Our results demonstrate that the likelihood and
efficiency of transformation into a dSph are both enhanced
substantially when the progenitor rotationally-supported dwarfs
possess shallow DM density profiles, whose slopes in the central
stellar regions are in agreement with those inferred from recent
theoretical and observational efforts in dwarf galaxy formation. These
findings further establish tidal stirring as a prevalent evolutionary
mechanism for shaping the nature of dwarf galaxies within the
currently favored cold dark matter (CDM) cosmological paradigm.

\section{Methods}
\label{section:methods}

A description of the adopted methodology is given in
\citet{Lokas_etal12a}. For completeness, we provide a summary of our
approach here. We employed the technique of \citet{Widrow_etal08} to
construct numerical realizations of dwarf galaxies consisting of
exponential stellar disks embedded in DM halos. The DM halo density
profiles followed the general form \citep{Lokas02},
\begin{equation}
  \rho(r) = 
  \frac{\rho_s} {(r/r_s)^\gamma \,
    (1 + r/r_s)^{3-\gamma}} \ ,
\end{equation}
where $\rho_s$, $r_s$, and $\gamma$ denote the characteristic inner
density, the scale radius, and the {\it asymptotic} inner slope of the
profile, respectively. $\rho_s$ depends on the epoch of halo
formation, the present-day values of the cosmological parameters, and
$\gamma$ (throughout the paper we assume the concordance {\LCDM}
cosmogony and $z=0$).

To examine how the DM density distribution in the central stellar
region of the progenitor disky dwarf affects tidal stirring, we varied
$\gamma$ in three otherwise identically initialized dwarf galaxies.
Specifically, we adopted $\gamma=1$ (corresponding to the NFW profile)
and two shallower inner slopes, namely a mild density cusp
($\gamma=0.6$) and a nearly constant density core ($\gamma=0.2$).
These shallow power-law indices are well-motivated as they are akin to
those of both observed dIrrs \citep[e.g.,][]{Weldrake_etal03,
  Oh_etal11} and realistic dIrr-like systems formed in hydrodynamical
cosmological simulations \citep{Governato_etal10}.

Each dwarf galaxy comprised a DM halo with a virial mass of $M_{\rm
  vir} = 10^{9} M_{\odot}$ (corresponding to a virial radius of
$r_{\rm vir} \approx 25.9$~kpc) and a concentration parameter of $c
\equiv r_{\rm vir}/r_s = 20$ ($r_s \approx 1.29$~kpc), and was
exponentially truncated beyond $r_{\rm vir}$
\citep{Kazantzidis_etal04a}. All DM halos hosted an identical stellar
disk whose mass, radial scale-length, sech$^2$ vertical scale-height,
and central radial velocity dispersion were equal to $M_d=0.02 M_{\rm
  vir}$, $R_d \approx 0.41$~kpc, $z_d =0.2 R_d$, and
$\sigma_{R0}=10\kms$, respectively (see \citealt{Kazantzidis_etal11a}
for the motivation behind these parameter values). We note that the
value of $R_d$ is derived assuming a dimensionless halo spin parameter
of $\lambda=0.04$ \citep{Mo_etal98}.  Being embedded in different DM
density distributions, the resulting disks differed in their velocity
dispersion profile and Toomre $Q$ stability parameter (at $R=2.5 R_d$,
$Q \approx 3.8$, $3.3$, and $2.9$ for $\gamma=1$, $0.6$ and $0.2$,
respectively; direct numerical simulations of the evolution of all
dwarf galaxies in isolation for a period of $10$~Gyr confirmed their
stability against bar formation).

Our choices of $\gamma$ above lead to DM density profiles, circular
velocity profiles, and stellar binding energy distributions that
display markedly dissimilar shapes within $5 R_d \approx 2$~kpc,
namely the radius containing the vast majority ($\gtrsim 95\%$) of
disk stars (see Figure~1 of \citealt{Lokas_etal12a}). Hence, although
our modeling approach is certainly not unique, it does satisfy the
main requirement of the present study by ensuring that the employed
rotationally-supported dwarfs exhibit substantially different initial
properties in the central stellar region. In particular, decreasing
cusp slopes correspond to less concentrated mass and energy
distributions, less steeply rising circular velocity profiles, and
smaller maximum circular velocities $V_{\rm max}$ ($V_{\rm max}
\approx 19.6$, $17.8$, and $16.7\kms$ for $\gamma=1$, $0.6$ and $0.2$,
respectively).  Therefore, rotationally-supported dwarfs embedded in
DM halos with shallow density profiles are expected to manifest rather
different responses to tidal shocks and overall tidal evolutions
compared to their counterparts with steeper DM density distributions.
We shall quantify these differences in the following section.


\begin{table}
\caption{Orbital Parameters of Disky Dwarfs}
\begin{center}
  \vspace*{-12pt}
\begin{tabular}{lcccc}
\hline
\hline
\\
\multicolumn{1}{c}{}                           &
\multicolumn{1}{c}{$r_{\rm apo}$}               &
\multicolumn{1}{c}{}                           &
\multicolumn{1}{c}{$r_{\rm peri}$}              &
\multicolumn{1}{c}{$T_{\rm orb}$}               
\\
\multicolumn{1}{c}{Orbit}                      &
\multicolumn{1}{c}{(kpc)}                      & 
\multicolumn{1}{c}{$r_{\rm apo}/r_{\rm peri}$}    &
\multicolumn{1}{c}{(kpc)}                      & 
\multicolumn{1}{c}{(Gyr)}                      
\\
\multicolumn{1}{c}{(1)}                        &
\multicolumn{1}{c}{(2)}                        &
\multicolumn{1}{c}{(3)}                        &
\multicolumn{1}{c}{(4)}                        &
\multicolumn{1}{c}{(5)}                        
\\
\\
\hline
\\
O1 &  125  &  5   &  25    &  2.1  \\
O2 &  85   &  5   &  17    &  1.3  \\
O3 &  250  &  5   &  50    &  5.4  \\
O4 &  125  &  10  &  12.5  &  1.8  \\
O5 &  125  &  2.5 &  50    &  2.5  \\
\hline
\end{tabular}
\end{center}
\vspace{-0.2cm}
\label{table:orbital_parameters}
\end{table}



\begin{figure*}
\begin{center}
\begin{tabular}{c}
  \includegraphics[scale=0.42]{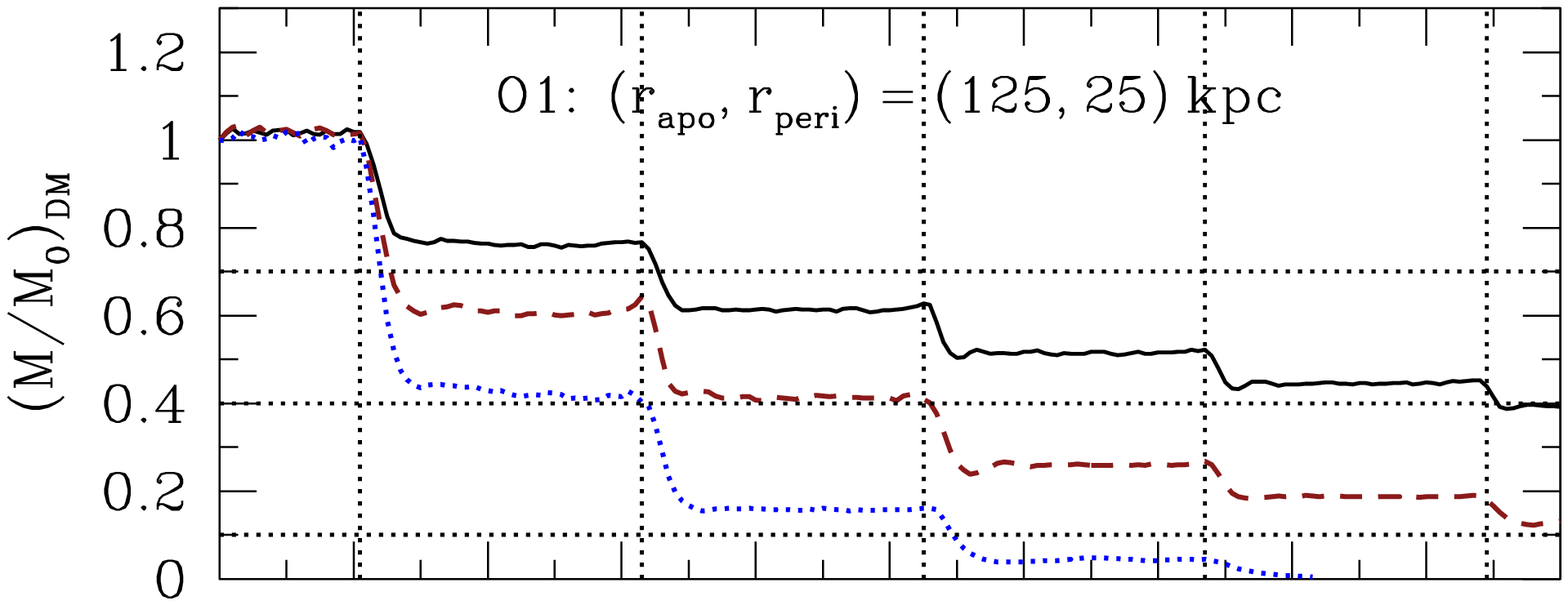}\\
  \includegraphics[scale=0.42]{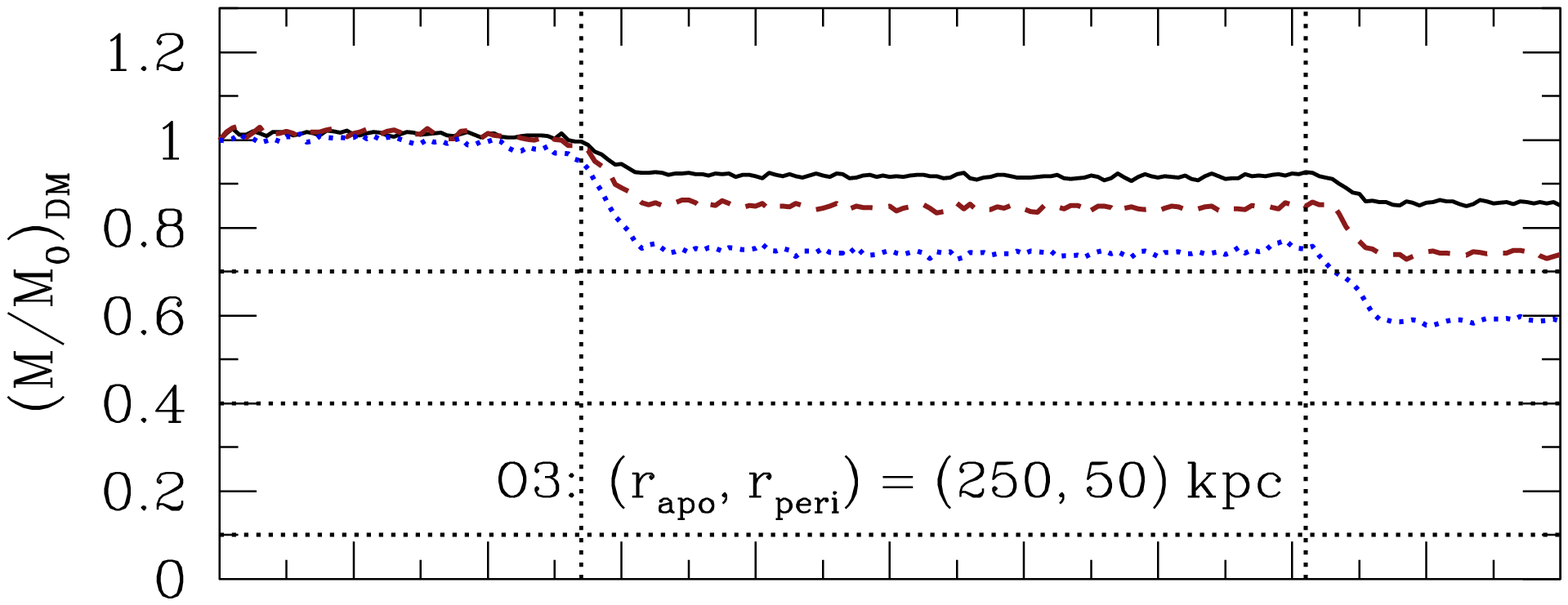}\\
  \includegraphics[scale=0.42]{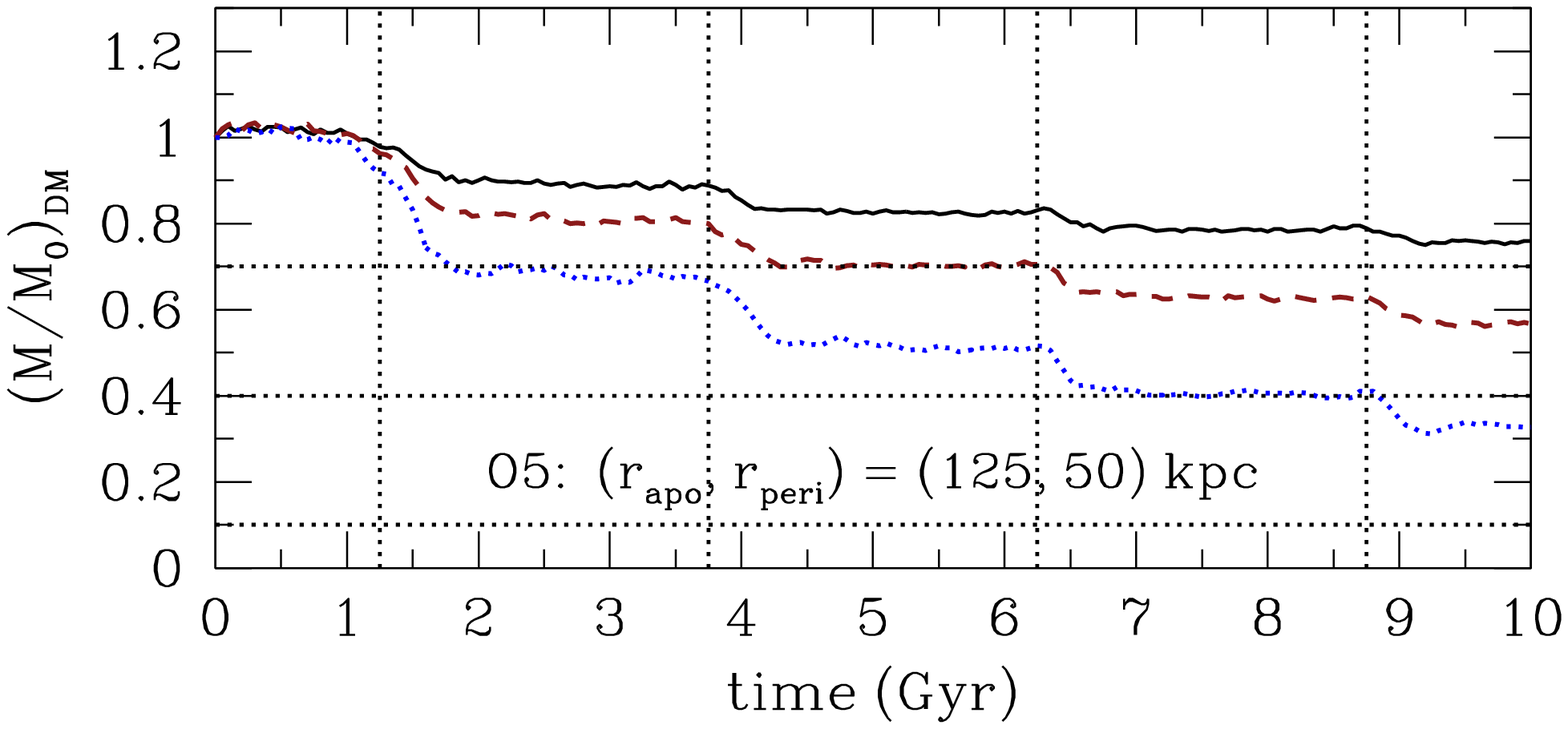}
\end{tabular}
\hspace{-0.5cm}
\begin{tabular}{c}
  \includegraphics[scale=0.42]{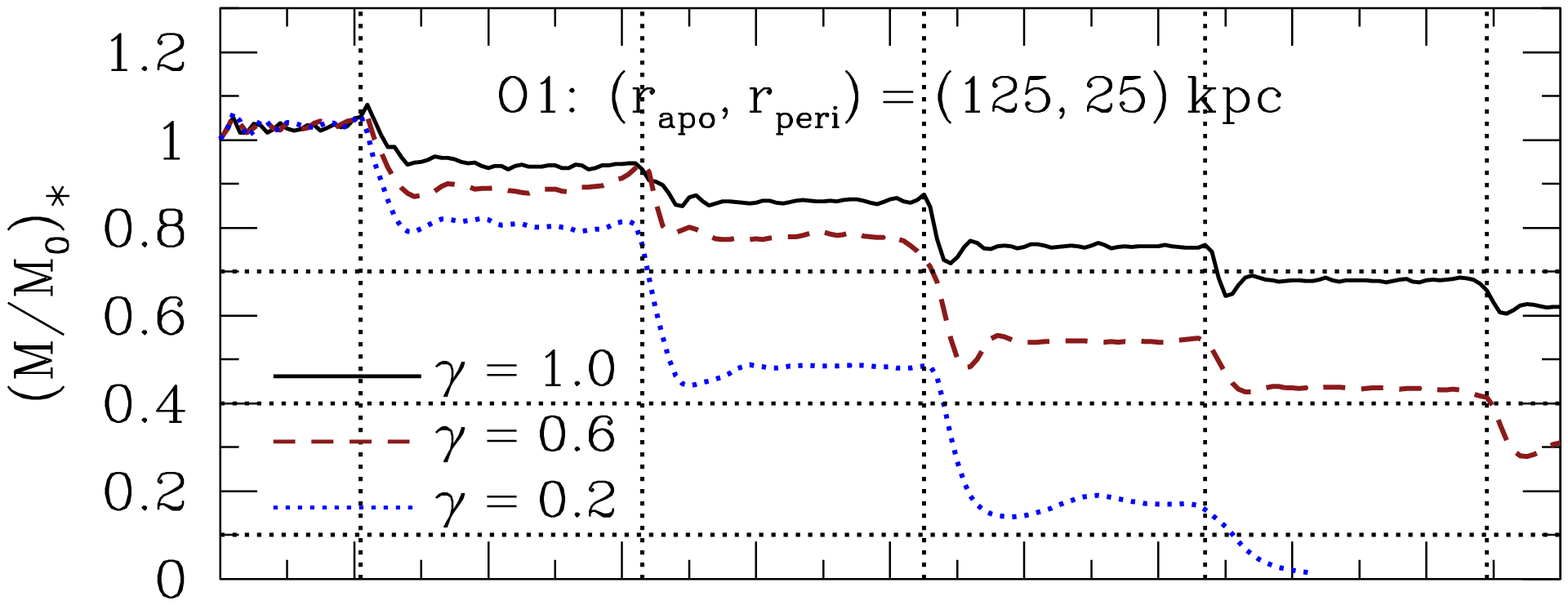}\\
  \includegraphics[scale=0.42]{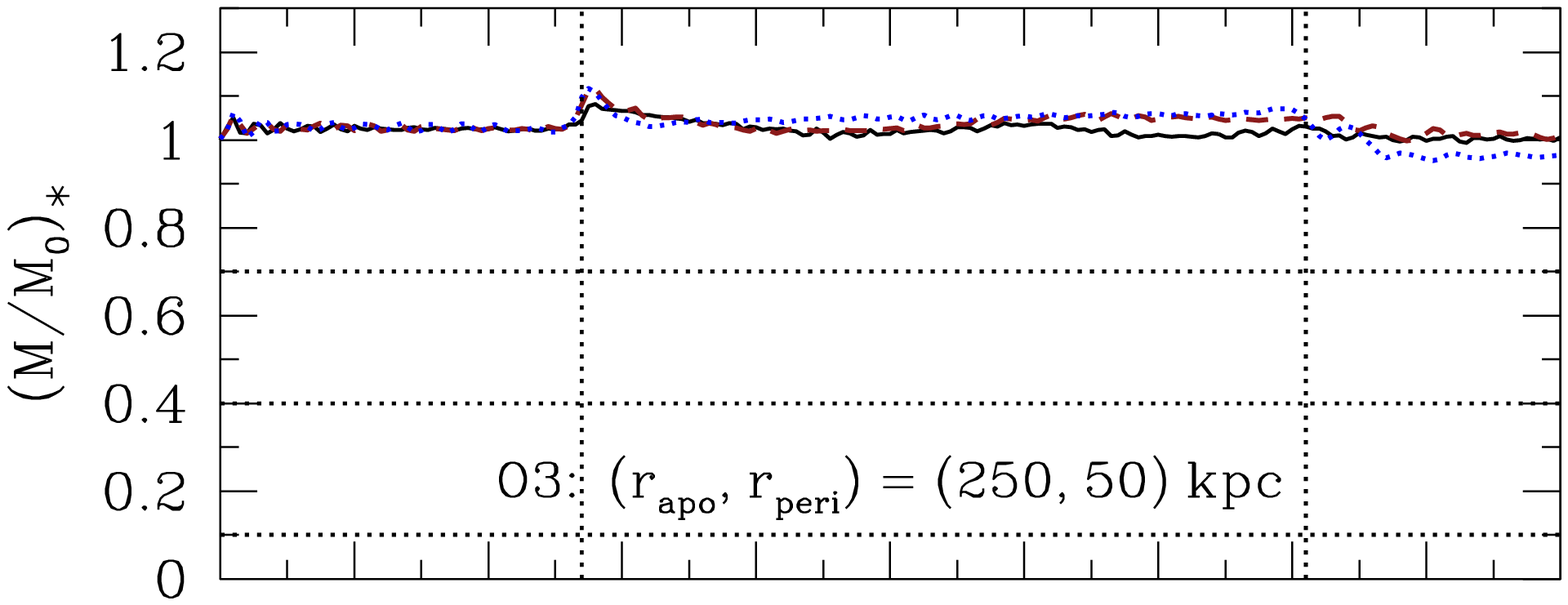}\\
  \includegraphics[scale=0.42]{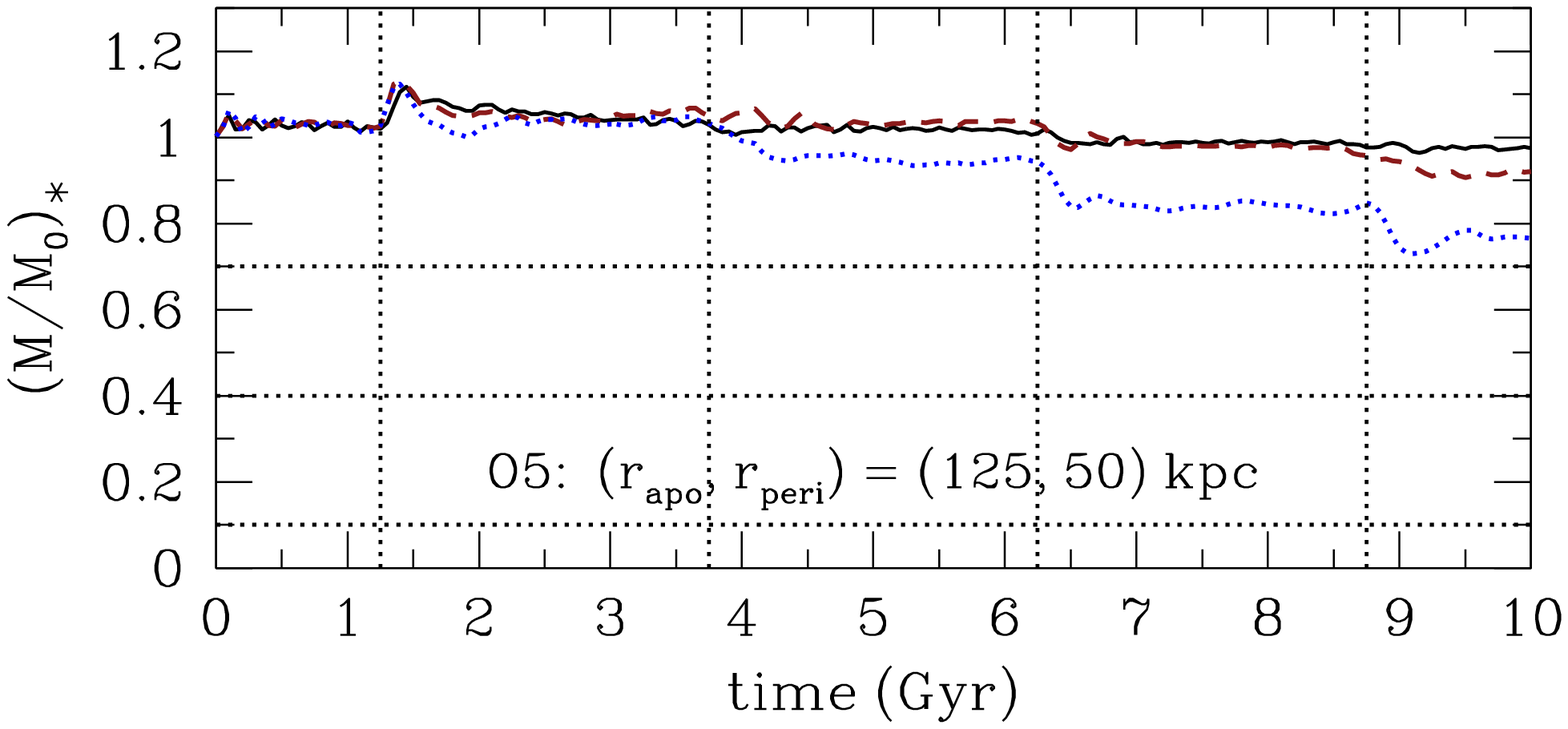}
\end{tabular}
\end{center}
\vspace{-0.5cm}
\caption{Evolution of mass in the DM (left panels) and stellar (right
  panels) components of the simulated disky dwarf galaxies as a
  function of time.  Results are presented for orbits O1 (upper
  panels), O3 (middle panels), and O5 (lower panels). $\gamma$ denotes
  the {\it asymptotic} inner slope of the dwarf DM density
  distribution and the solid, dashed, and dotted lines correspond to
  $\gamma = 1.0$, $\gamma = 0.6$, and $\gamma = 0.2$, respectively. DM
  (stellar) masses are computed within $0.7$~kpc from the center of
  the dwarf (see text for details) and are normalized to the {\it
    initial} DM (stellar) mass enclosed within $0.7$~kpc, $M_0$. In
  all panels, vertical lines specify pericentric passages and
  horizontal lines indicate mass loss of $30\%$, $60\%$, and $90\%$
  with respect to the initial value. For a given orbit inside the host
  galaxy, rotationally-supported dwarfs embedded in DM halos with
  shallow density profiles ($\gamma<1$) suffer enhanced mass loss
  compared to their counterparts with steeper DM density distributions
  ($\gamma=1$).
  \label{fig1}}
\end{figure*}


The dwarf galaxy models contained $N_h = 10^6$ DM and $N_d = 5 \times
10^5$ disk particles with a gravitational softening of
$\epsilon_h=60$~pc and $\epsilon_d=20$~pc, respectively. Resolution
was adequate to resolve all scales of interest. We assumed a single
host represented by a self-gravitating, high-resolution numerical
model of the MW \citep{Kazantzidis_etal11a}.  Each disky dwarf was
placed on five bound orbits of varying sizes and eccentricities inside
the primary galaxy, for a total of $15$ experiments. In detail, we
employed a subset of orbits from \citet{Kazantzidis_etal11a}, whose
parameters were motivated by both theoretical studies of the orbital
distributions of cosmological halos in MW-sized hosts
\citep[e.g.,][]{Diemand_etal07,Klimentowski_etal10} and observational
work pertaining to LG dwarfs \citep[e.g.,][]{McConnachie12}.
Table~\ref{table:orbital_parameters} summarizes the adopted orbital
parameters, including apocentric distances, $r_{\rm apo}$,
eccentricities, $r_{\rm apo}/r_{\rm peri}$, pericentric distances,
$r_{\rm peri}$, and orbital times, $T_{\rm orb}$. The dwarf galaxies,
having considerably lower masses with respect to the host, do not
experience dynamical friction \citep[e.g.,][]{Colpi_etal99} and, as a
result, their orbital parameters remain virtually unchanged during the
course of their evolution. In all cases, the dwarfs started at
apocenter and were evolved for $10$~Gyr inside the primary.  The
alignments between the internal angular momenta of the dwarfs, those
of the primary disks and the orbital angular momenta were all always
prograde and equal to $45^{\circ}$ (see \citealt{Kazantzidis_etal11a}
regarding the implications of these choices).  Lastly, the simulations
were performed with the $N$-body code PKDGRAV \citep{Stadel01}.

\section{Results}
\label{sec:results}

We examined the global response of the disky dwarfs subject to the
tidal field of the host galaxy via the time evolution of their masses,
kinematics, and shapes. These properties were always computed within
$0.7$~kpc from the dwarf center, a radius which is approximately equal
to the half-light radius, $r_{1/2}$, of the initial disk. Adopting
this well-defined, fixed scale facilitates meaningful comparisons
among different experiments and enables us to overcome the
complications associated with determining tidal radii
\citep[e.g.,][]{Read_etal06}.

The dwarf kinematics and shapes were quantified through the parameters
$V_{\rm rot}/\sigma_{\ast}$ and $c/a$, where $V_{\rm rot}$,
$\sigma_{\ast}$, $c$, and $a$ denote the rotational velocity, the
one-dimensional velocity dispersion, the minor axis, and the major
axis of the stellar distribution, respectively. At each simulation
output, the following procedure was repeated. First, we determine the
directions of the principal axes and derive $c/a$ using the moments of
the inertia tensor. Subsequently, we introduce a spherical coordinate
system ($r$, $\theta$, $\phi$), where $\phi$ is the angle coordinate
in the $xy$ plane oriented in such a way that the $z$-axis is along
the minor axis of the stellar distribution.  Lastly, we calculate
$V_{\rm rot}$ around the minor axis $V_{\rm rot}=V_{\phi}$ and compute
the dispersions $\sigma_r$, $\sigma_{\theta}$, and $\sigma_{\phi}$
around the mean values; $\sigma_{\ast}$, which we adopt throughout the
paper as a measure of the amount of random motions in the stars, is
defined as $\sigma_{\ast} \equiv [(\sigma_r^2 + \sigma_{\theta}^2 +
\sigma_{\phi}^2)/3]^{1/2}$.


\begin{table*}
\caption{Summary of Results}
\begin{center}
  \vspace*{-12pt}
\begin{tabular}{lccccccccc}
\hline
\hline
\\
\multicolumn{1}{c}{}                           &
\multicolumn{1}{c}{}                           &
\multicolumn{1}{c}{}                           &
\multicolumn{1}{c}{}                           &
\multicolumn{1}{c}{$t_d$}                      &
\multicolumn{1}{c}{Bar}                        &
\multicolumn{1}{c}{}                           &
\multicolumn{1}{c}{}                           &
\multicolumn{1}{c}{}                           &
\multicolumn{1}{c}{$t_{\rm dSph}$}              
\\
\multicolumn{1}{c}{Simulation}                 & 
\multicolumn{1}{c}{$\gamma$}                   & 
\multicolumn{1}{c}{Orbit}                      &
\multicolumn{1}{c}{Survival}                   &
\multicolumn{1}{c}{(Gyr)}                      &
\multicolumn{1}{c}{Formation}                  &
\multicolumn{1}{c}{$V_{\rm rot}/\sigma_{\ast}$}  & 
\multicolumn{1}{c}{$c/a$}                      &
\multicolumn{1}{c}{Classification}             &
\multicolumn{1}{c}{(Gyr)}                      
\\
\multicolumn{1}{c}{(1)}                        &
\multicolumn{1}{c}{(2)}                        &
\multicolumn{1}{c}{(3)}                        &
\multicolumn{1}{c}{(4)}                        &
\multicolumn{1}{c}{(5)}                        &
\multicolumn{1}{c}{(6)}                        &
\multicolumn{1}{c}{(7)}                        &
\multicolumn{1}{c}{(8)}                        &
\multicolumn{1}{c}{(9)}                        &
\multicolumn{1}{c}{(10)}                       
\\
\\
\hline
\\
S1   & 1.0  &  O1 & yes & $-$   & yes & 0.63 & 0.66 & ~dSph?    & $-$       \\
S2   & 1.0  &  O2 & yes & $-$   & yes & 0.01 & 0.93 & dSph      & 2.85 (2)  \\
S3   & 1.0  &  O3 & yes & $-$   & no  & 1.21 & 0.33 & non-dSph  & $-$       \\
S4   & 1.0  &  O4 & yes & $-$   & yes & 0.06 & 0.93 & dSph      & 3.75 (2)  \\
S5   & 1.0  &  O5 & yes & $-$   & no  & 1.11 & 0.38 & non-dSph  & $-$       \\
\hline
\vspace*{-0.2cm}
\\
S6   & 0.6  &  O1 & yes & $-$   & yes & 0.01 & 0.82 & dSph      & 5.55 (3)   \\
S7   & 0.6  &  O2 & no  & 6.40  & yes & 0.01 & 0.82 & dSph      & 2.55 (2)   \\
S8   & 0.6  &  O3 & yes & $-$   & no  & 0.92 & 0.40 & non-dSph  & $-$        \\
S9   & 0.6  &  O4 & no  & 9.30  & yes & 0.07 & 0.76 & dSph      & 3.35 (2)   \\
S10  & 0.6  &  O5 & yes & $-$   & yes & 0.67 & 0.49 & ~dSph?    & $-$        \\
\hline                                              
\vspace*{-0.2cm}                                           
\\
S11  & 0.2  &  O1 & no  & 8.15  & yes & 0.06 & 0.83 & dSph      & 3.70 (2)   \\
S12  & 0.2  &  O2 & no  & 3.80  & yes & 0.07 & 0.82 & dSph      & 1.85 (1)   \\
S13  & 0.2  &  O3 & yes & $-$   & yes & 0.63 & 0.47 & ~dSph?    & $-$        \\
S14  & 0.2  &  O4 & no  & 3.35  & yes & 0.15 & 0.68 & dSph      & 2.60 (1)   \\
S15  & 0.2  &  O5 & yes & $-$   & yes & 0.64 & 0.56 & ~dSph?    & $-$        \\
\hline
\end{tabular}
\end{center}
\label{table:summary}
\end{table*}


Given the typical parameters associated with orbit O1, we use it as
the basis for the comparison with the other experiments. In what
follows, we only describe results pertaining to orbits O1, O3, and O5.
This is for brevity and because the evolution of the
rotationally-supported dwarfs on the tightest (O2) and the most
eccentric (O4) of our orbits displays similar salient features to
those of O1. We emphasize that orbits O3 and O5 are characterized by
larger pericenters and/or lower eccentricities compared to those of
representative CDM satellites \citep[e.g.,][]{Diemand_etal07}.

Figure~\ref{fig1} shows the time evolution of both DM and stellar mass
of the simulated disky dwarf galaxies as they orbit inside the
primary. Apart from illustrating the continuous stripping of the dwarf
mass by the host tidal field, this figure highlights a number of
additional generic features. First, masses decrease significantly at
pericentric passages, where the intensity and variation of the
time-dependent tidal forces are the strongest, and the tidal shocks
occur. Between pericenters, masses remain remarkably constant,
indicating that the dwarfs respond nearly adiabatically to the tidal
field.  Furthermore, the mass loss of the DM component greatly exceeds
that of the stellar distribution, reflecting the more efficient tidal
stripping of the {\it extended} DM halos of the dwarfs.

Second, irrespective of orbit, disky dwarf galaxies embedded in DM
halos with shallow density profiles ($\gamma<1$) suffer augmented mass
loss in both stellar and DM components compared to their counterparts
with steeper DM density distributions ($\gamma=1$) (see
\citealt{Kazantzidis_etal04b,Penarrubia_etal10} for similar
conclusions; these studies, however, performed simulations of the
tidal evolution of dwarfs without stellar disks). In some
circumstances, the amount of mass loss is so dramatic that it leads to
complete tidal disruption. Columns 4 and 5 of
Table~\ref{table:summary} provide information regarding dwarf survival
(as indicated by the presence of a self-bound entity at $t=10$~Gyr)
and list the time of disruption if it occurs, respectively.

\begin{figure*}[t]
\begin{center}
\begin{tabular}{c}
  \includegraphics[scale=0.42]{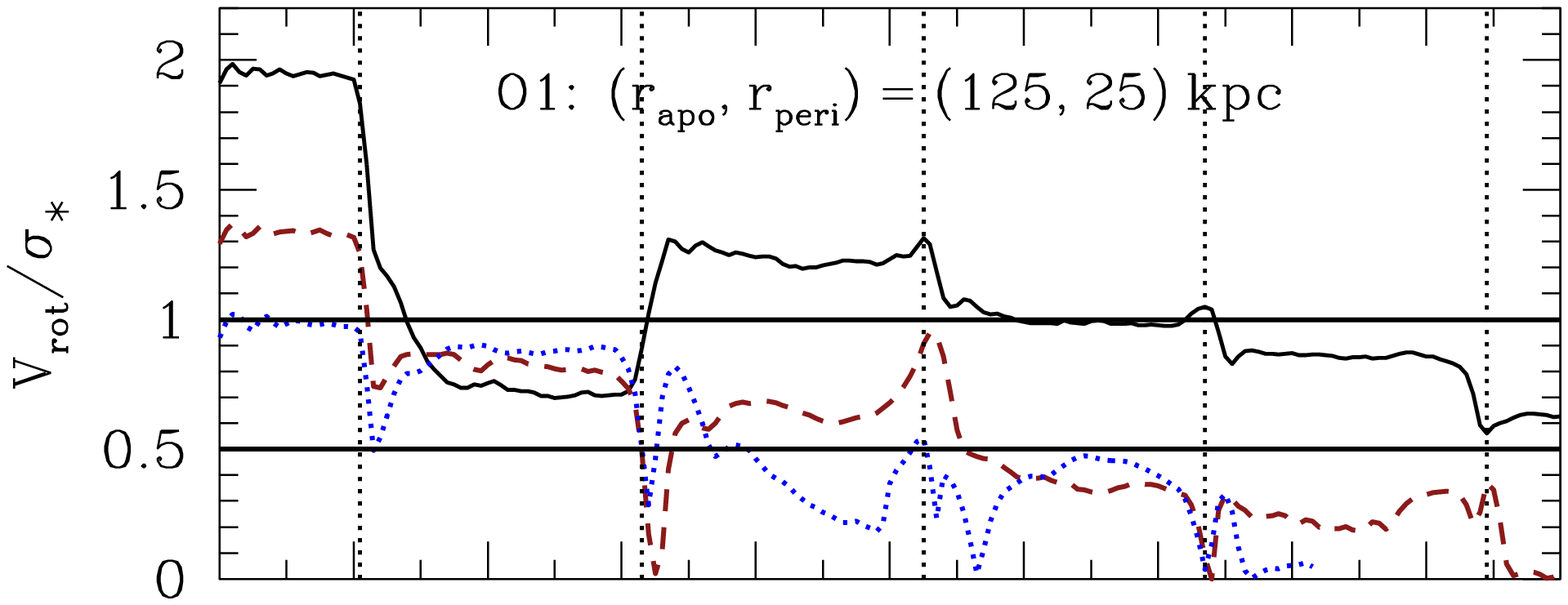}\\
  \includegraphics[scale=0.42]{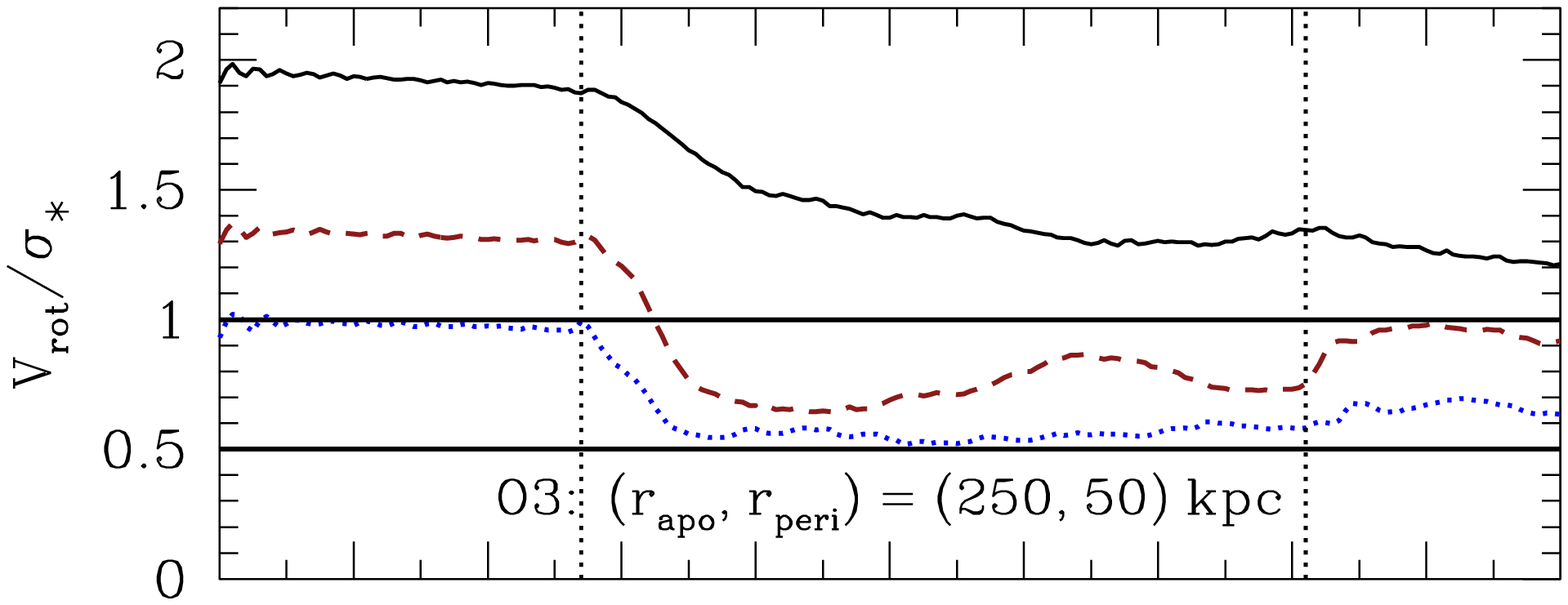}\\
  \includegraphics[scale=0.42]{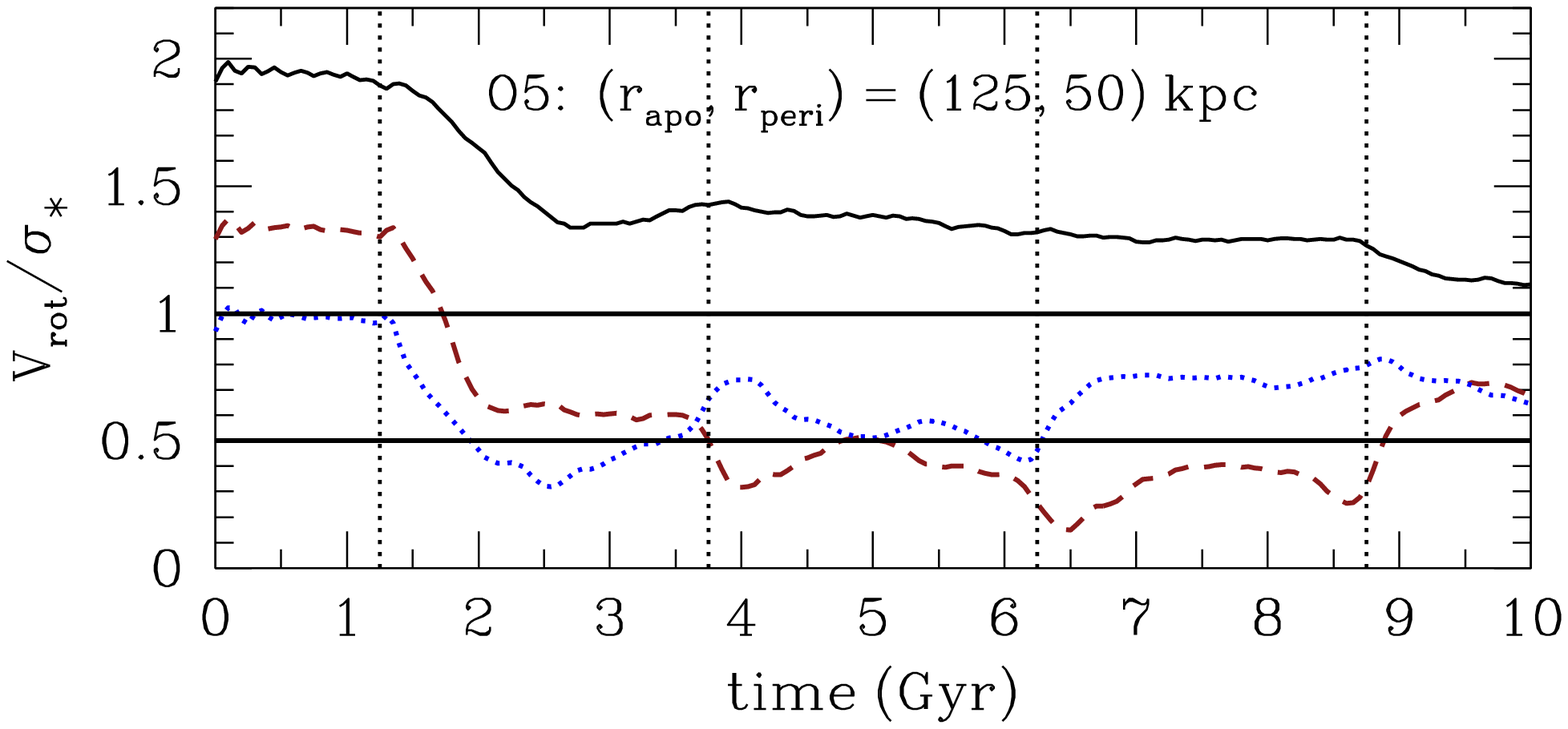}
\end{tabular}
\hspace{-0.5cm}
\begin{tabular}{c}
  \includegraphics[scale=0.42]{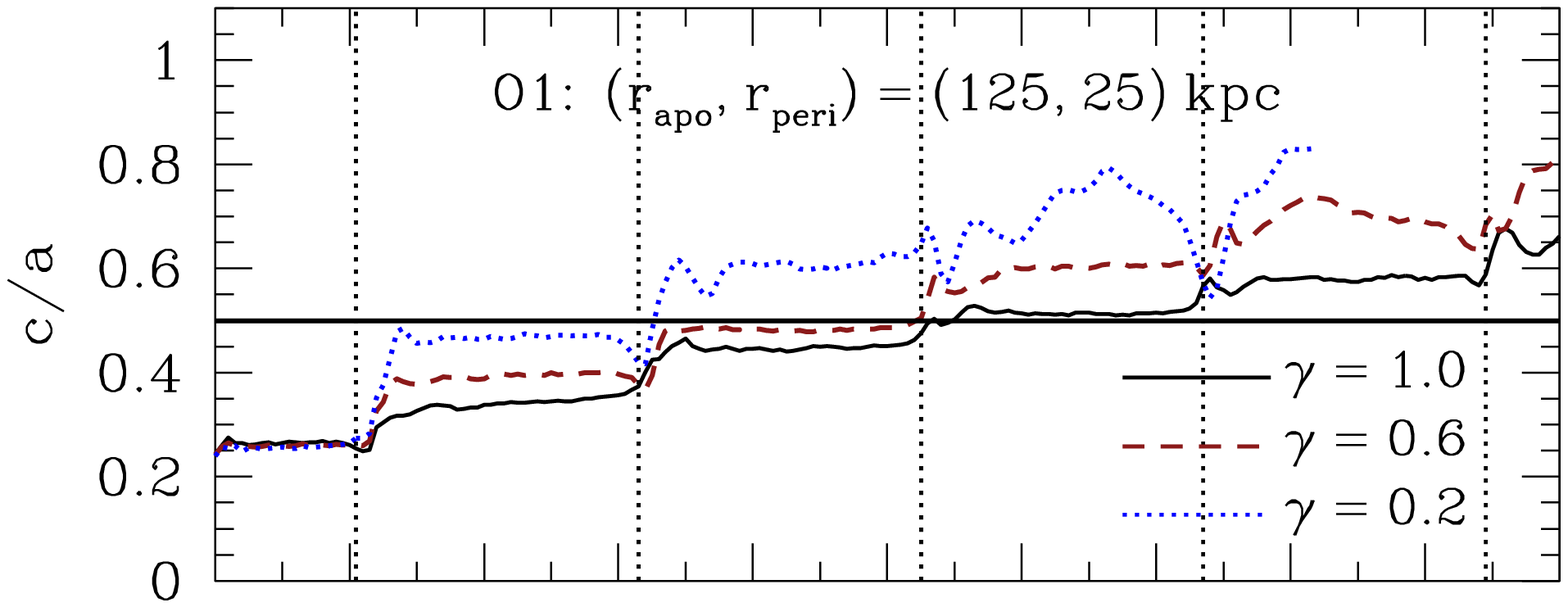}\\
  \includegraphics[scale=0.42]{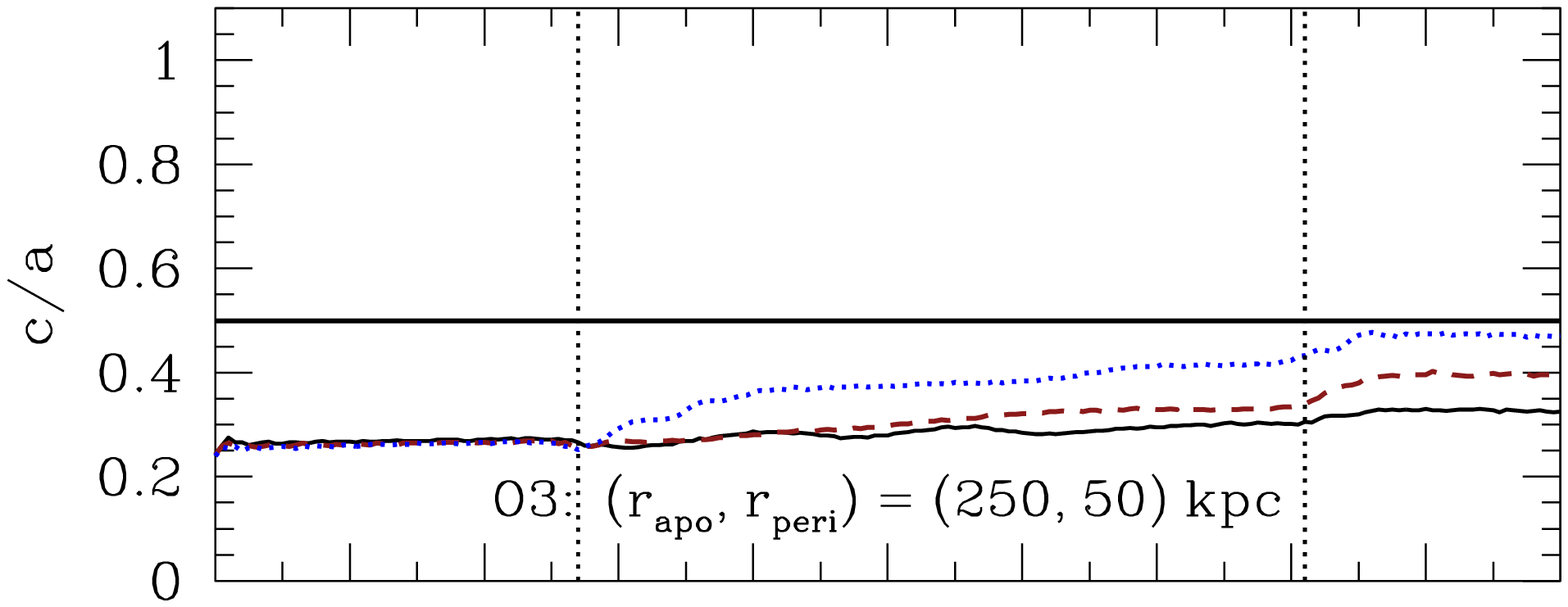}\\
  \includegraphics[scale=0.42]{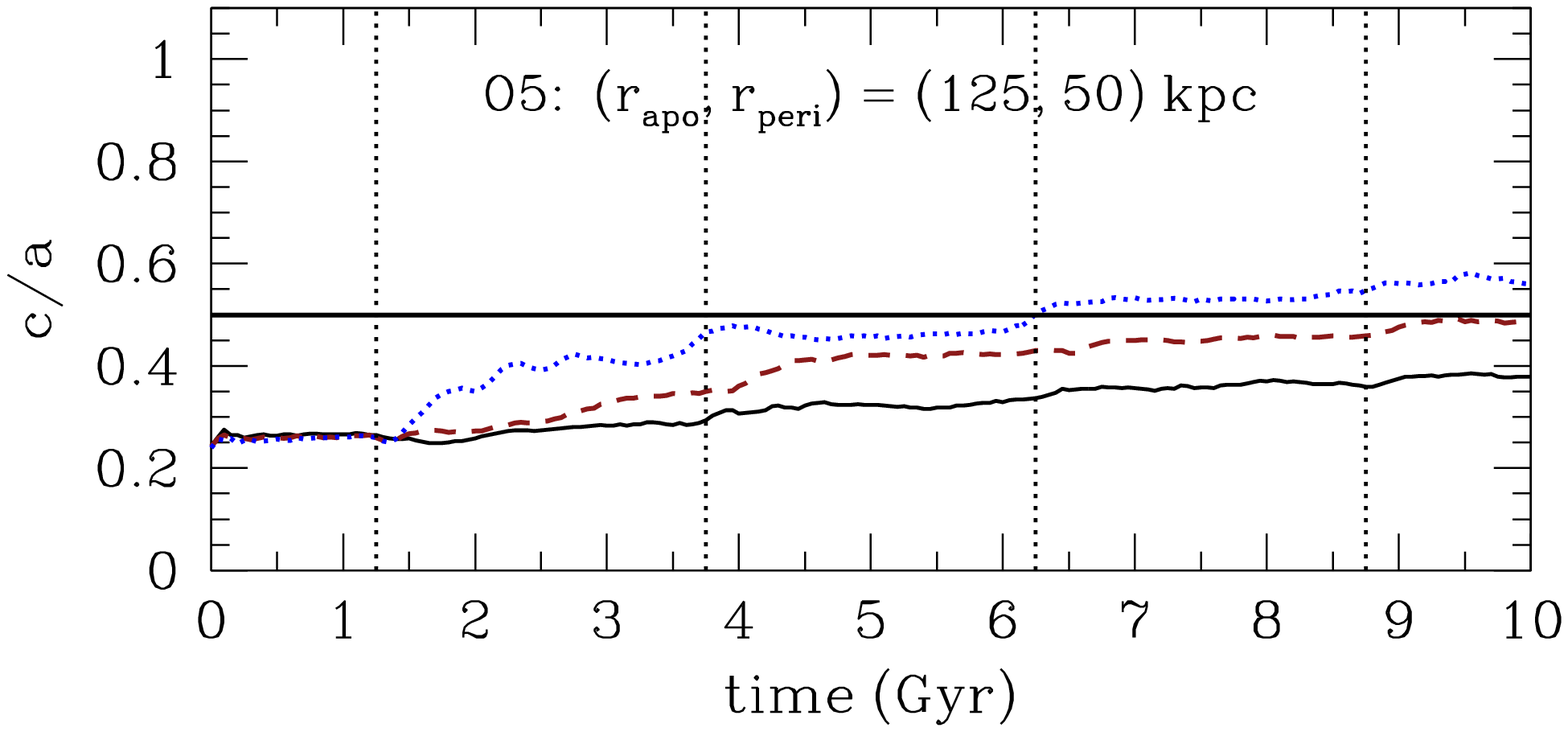}
\end{tabular}
\end{center}
\vspace{-0.5cm}
\caption{Evolution of stellar kinematics (left panels) and shape
  (right panels) of the simulated disky dwarf galaxies as a function
  of time.  Results are presented for orbits O1 (upper panels), O3
  (middle panels), and O5 (lower panels). Line types are as in
  Figure~\ref{fig1}.  $V_{\rm rot}/\sigma_{\ast}$ and $c/a$ are
  computed within $0.7$~kpc from the center of the dwarf (see text for
  details). Horizontal lines indicate the limiting values $V_{\rm
    rot}/\sigma_{\ast} =1$, $V_{\rm rot}/\sigma_{\ast} =0.5$, and
  $c/a=0.5$: simulated dwarfs whose final states are characterized by
  $V_{\rm rot}/\sigma_{\ast} \lesssim 0.5$ and $c/a \gtrsim 0.5$
  correspond to {\it bona fide} dSphs, while those with final values
  of $0.5 \lesssim V_{\rm rot}/\sigma_{\ast} \lesssim 1$ and $c/a
  \gtrsim 0.5$ are classified as ``dSph?'' (see text). For a given
  orbit inside the host, rotationally-supported dwarf galaxies embedded
  in DM halos with shallow density profiles ($\gamma<1$) experience a
  stronger evolution in their shapes and kinematics and demonstrate a
  considerably enhanced likelihood and efficiency of transformation 
  into dSph-like systems relative to their counterparts with steeper
  DM density distributions ($\gamma=1$).
  \label{fig2}}
\end{figure*}

In Figure~\ref{fig2}, we present the time evolution of the kinematics
and shapes of the simulated disky dwarfs as they orbit inside the host
galaxy.  Given that $V_{\rm rot}$ is larger for more concentrated mass
distributions and that all dwarf models are initialized with the same
value of $\sigma_{R0}$, steeper cusp slopes result in greater initial
values of $V_{\rm rot}/\sigma_{\ast}$. We also note that the
$\gamma=0.2$ dwarf exhibits $V_{\rm rot}/\sigma_{\ast} \approx 1$
initially. This notably low value of $V_{\rm rot}/\sigma_{\ast}$ for a
disk is simply a consequence of our specific choice to compute the
dwarf properties within $0.7$~kpc from the center and does not
interfere with the interpretation of the results.

Injection of energy via tidal shocks at pericentric passages increases
the random motions of disk stars, causing their orbits to become more
isotropic. Thus, $V_{\rm rot}/\sigma_{\ast}$ progressively decreases,
signifying the transition from rotationally-supported to
pressure-supported stellar systems dominated by random motions.
Simultaneously, the initially disky stellar distributions evolve into
more spheroidal shapes as reflected in the continuous increase of
$c/a$. Figure~\ref{fig2} shows that, regardless of orbit inside the
primary, rotationally-supported dwarfs embedded in DM halos with
shallow density profiles ($\gamma<1$) demonstrate stronger evolution
in their kinematics and shapes compared to their counterparts with
steeper DM density distributions ($\gamma=1$).

It is important to stress that the strong tidal forces at pericenters
typically trigger bar instabilities in the disks of the dwarfs.  These
tidally-induced bars play a vital role in the overall decrease of
$V_{\rm rot}$ by transporting angular momentum outwards.  As tidal
stripping removes the outer parts of the dwarfs, the entire angular
momentum content progressively decreases and the ability of the dwarf
galaxies to be supported by rotation progressively diminishes.  The
seemingly unexpected increase of $V_{\rm rot}/\sigma_{\ast}$ observed
in Figure~\ref{fig2} between some pericentric passages is due to the
particular orientation of these bars at the moment of pericenter
crossing and the strong tidal torques exerted on them by the primary
galaxy, resulting in an increase of $V_{\rm rot}$ (see
\citealt{Kazantzidis_etal11a}). Column 6 of Table~\ref{table:summary}
indicates that bar formation is more probable in shallow DM density
distributions (see also \citealt{Mayer_Wadsley04}).


\begin{figure*}[t]
\begin{center}
\begin{tabular}{c}
  \includegraphics[scale=0.35]{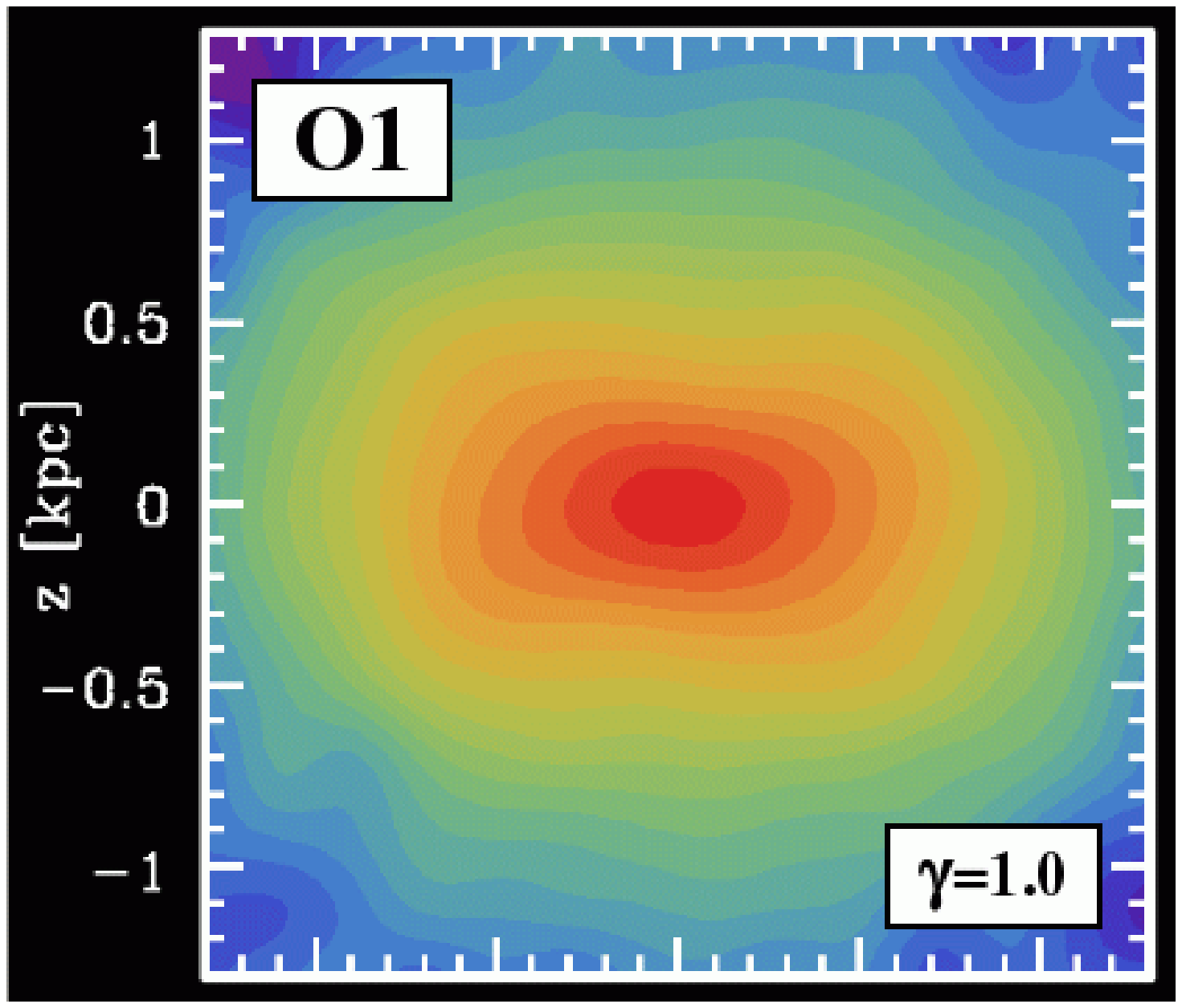}\hspace{-0.1cm}
  \includegraphics[scale=0.35]{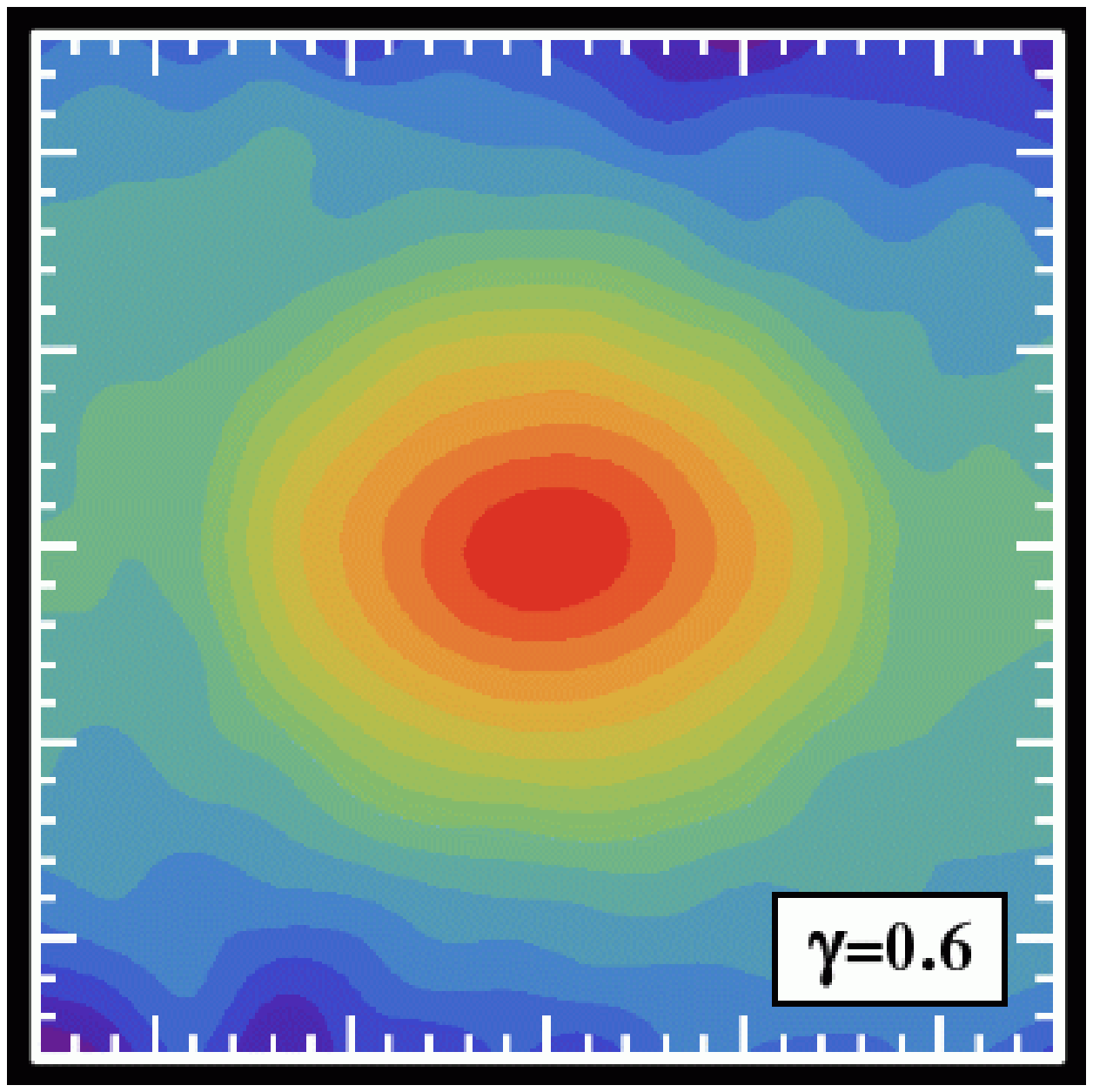}\hspace{-0.1cm}
  \includegraphics[scale=0.35]{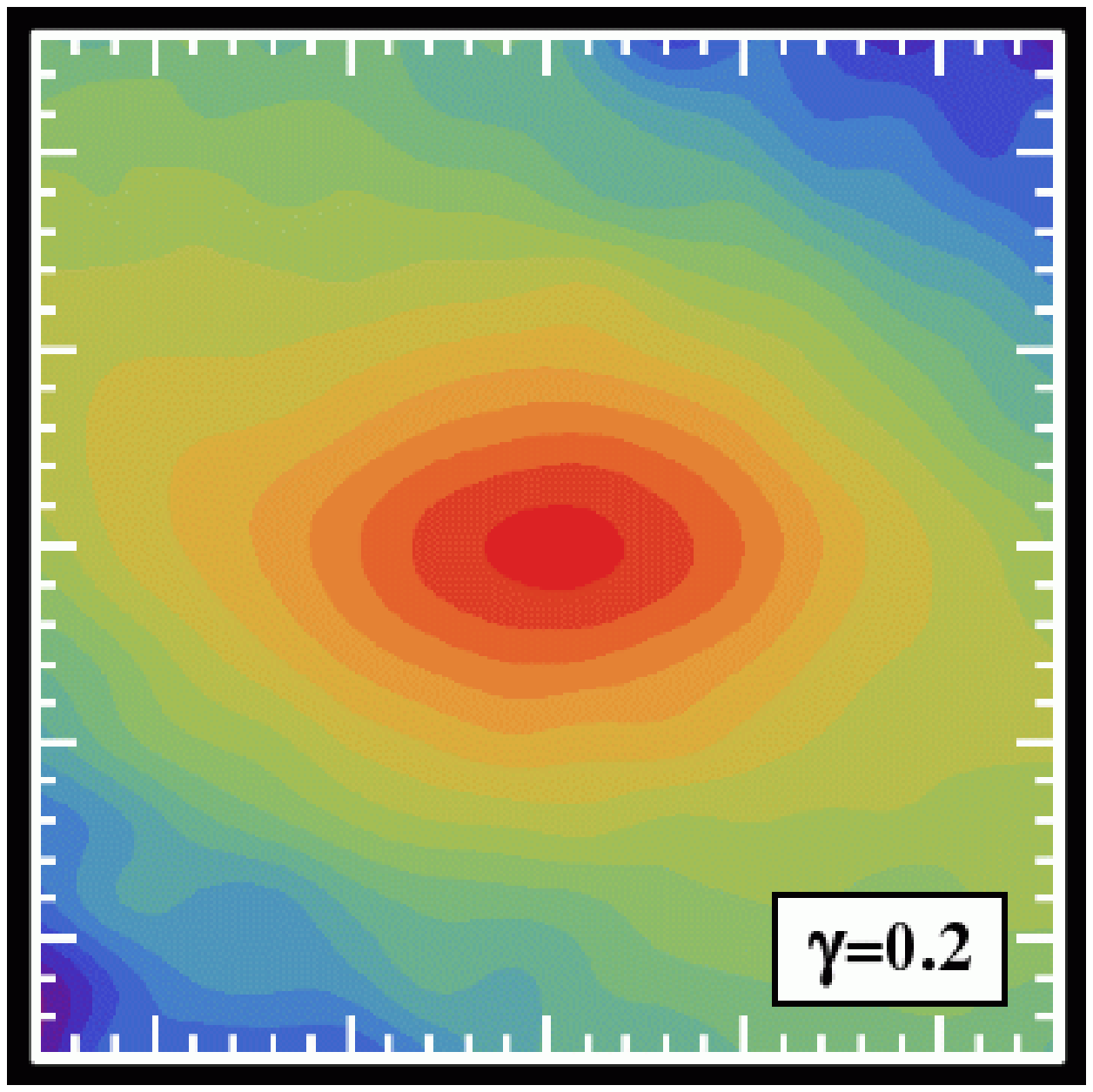}
  \vspace{-0.15cm}
\end{tabular}
\begin{tabular}{c}
  \includegraphics[scale=0.35]{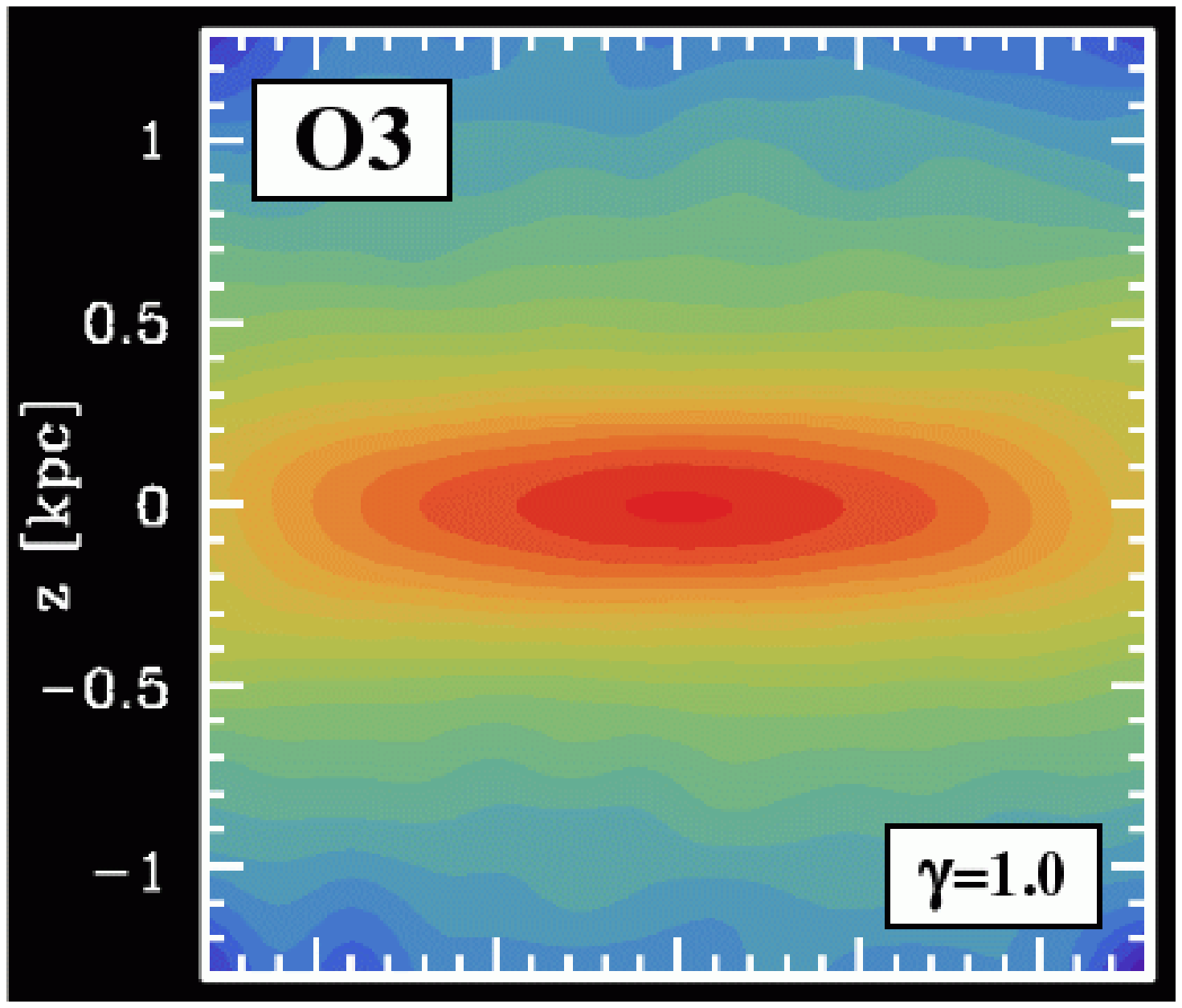}\hspace{-0.1cm}
  \includegraphics[scale=0.35]{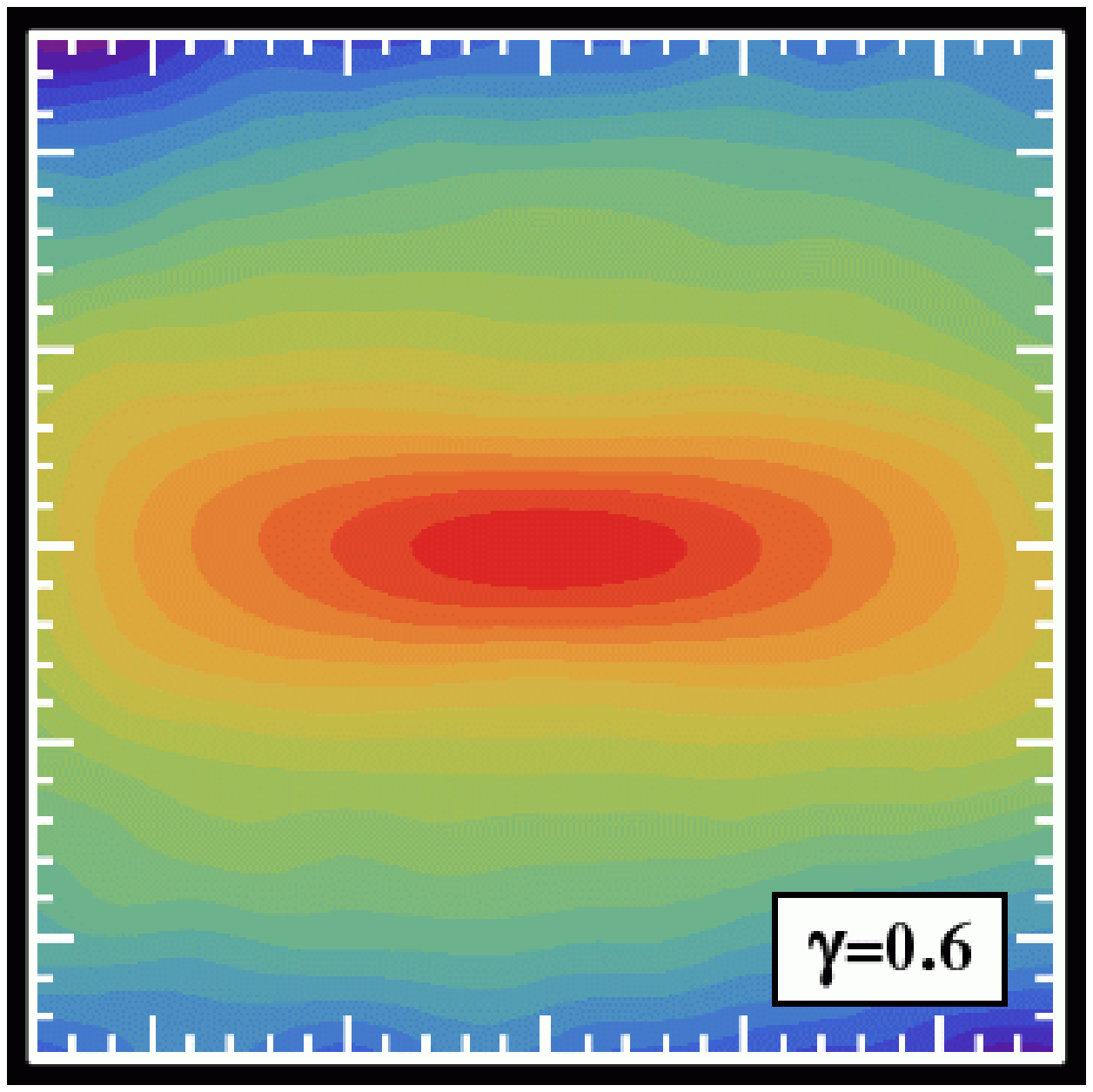}\hspace{-0.1cm}
  \includegraphics[scale=0.35]{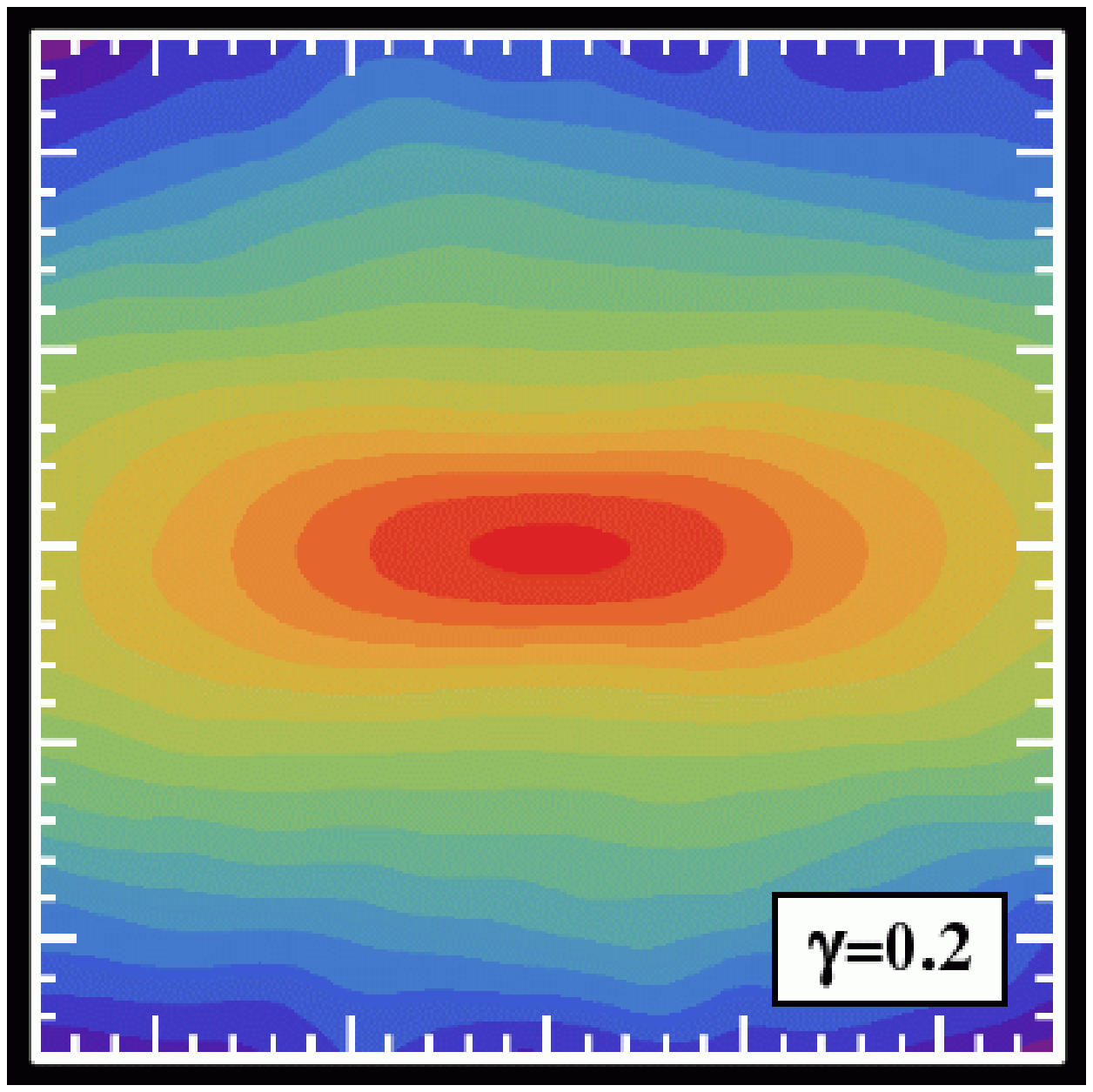}
  \vspace{-0.15cm}
\end{tabular}
\begin{tabular}{c}
  \includegraphics[scale=0.35]{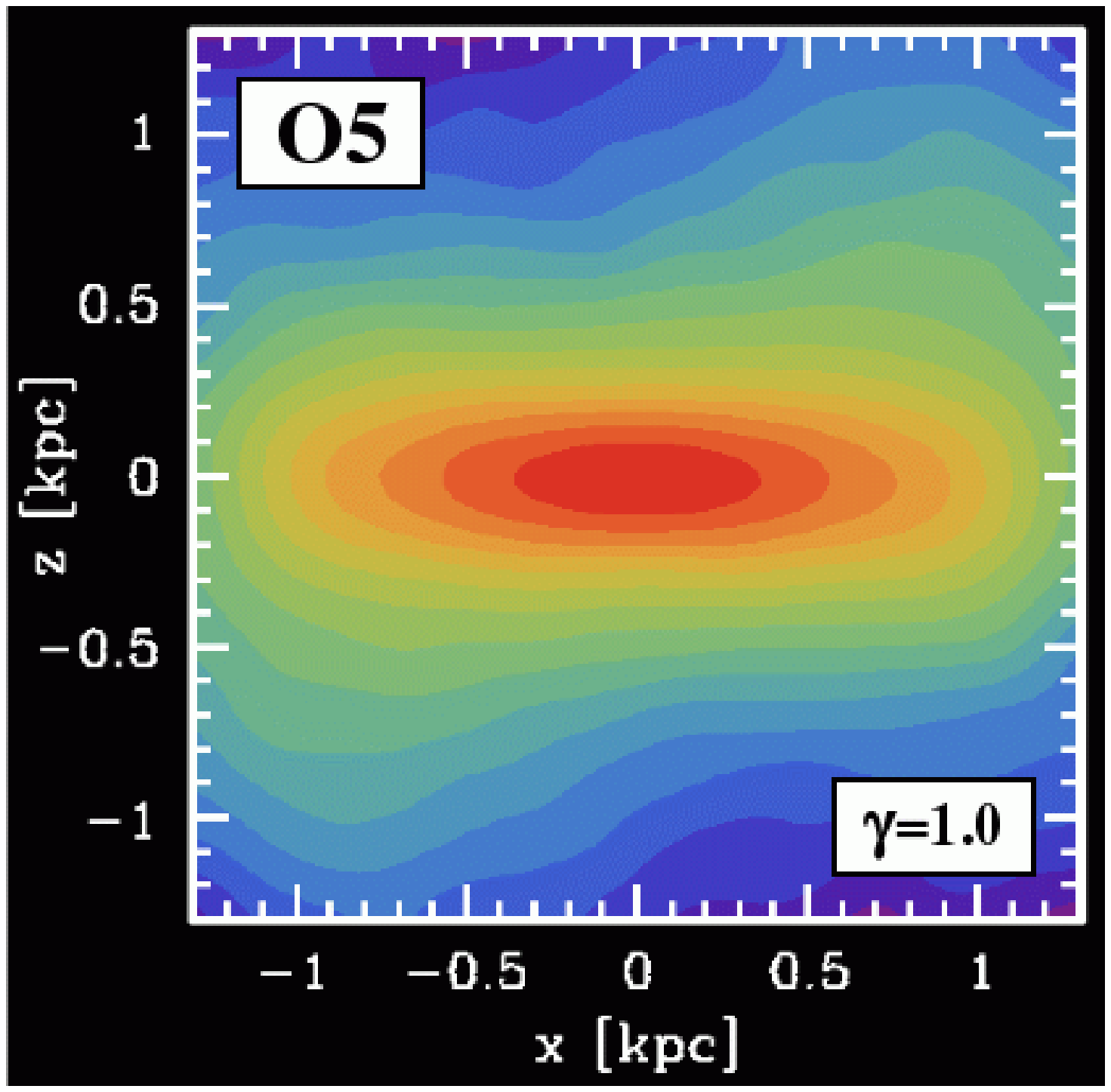}\hspace{-0.1cm}
  \includegraphics[scale=0.35]{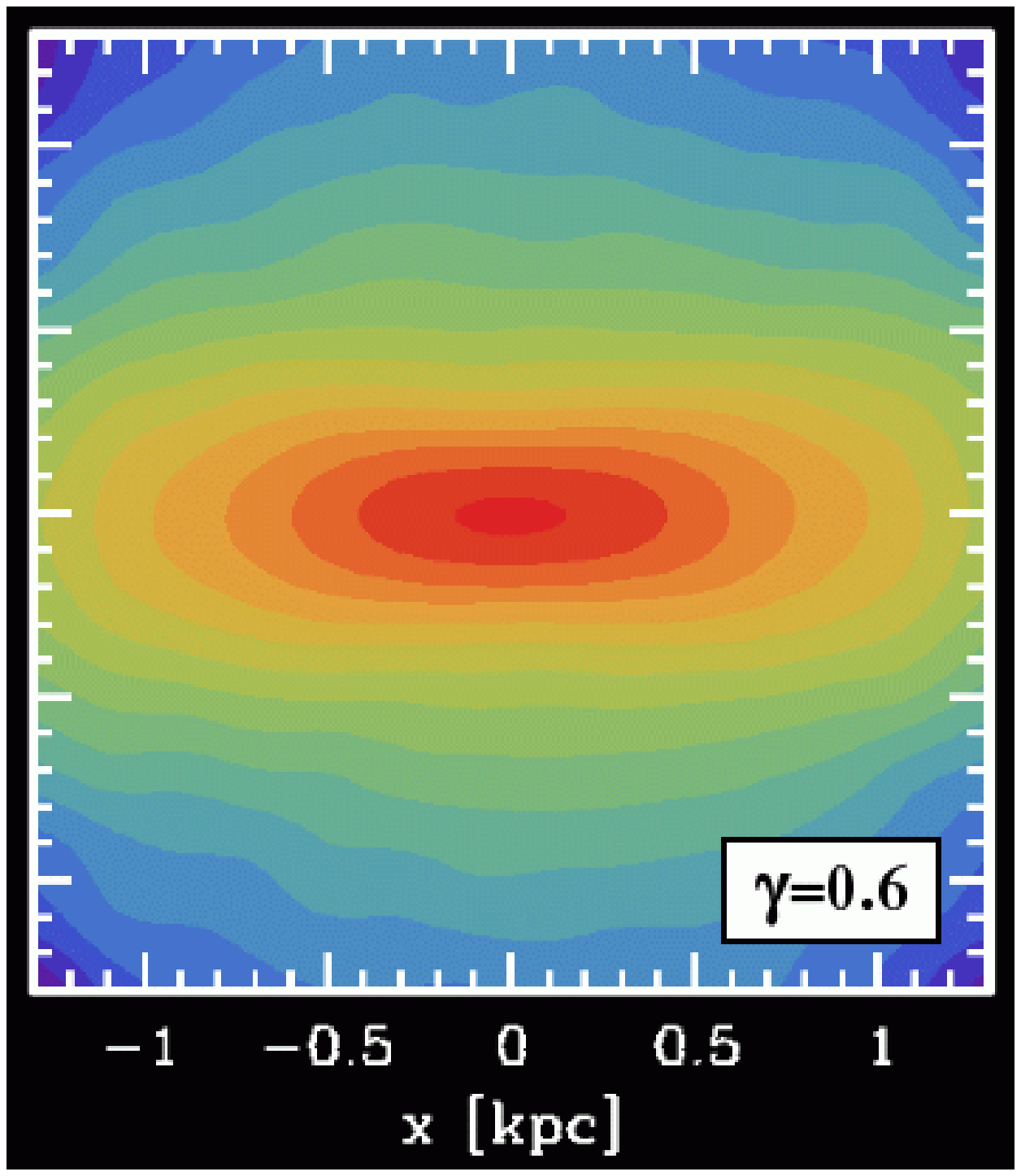}\hspace{-0.1cm}
  \includegraphics[scale=0.35]{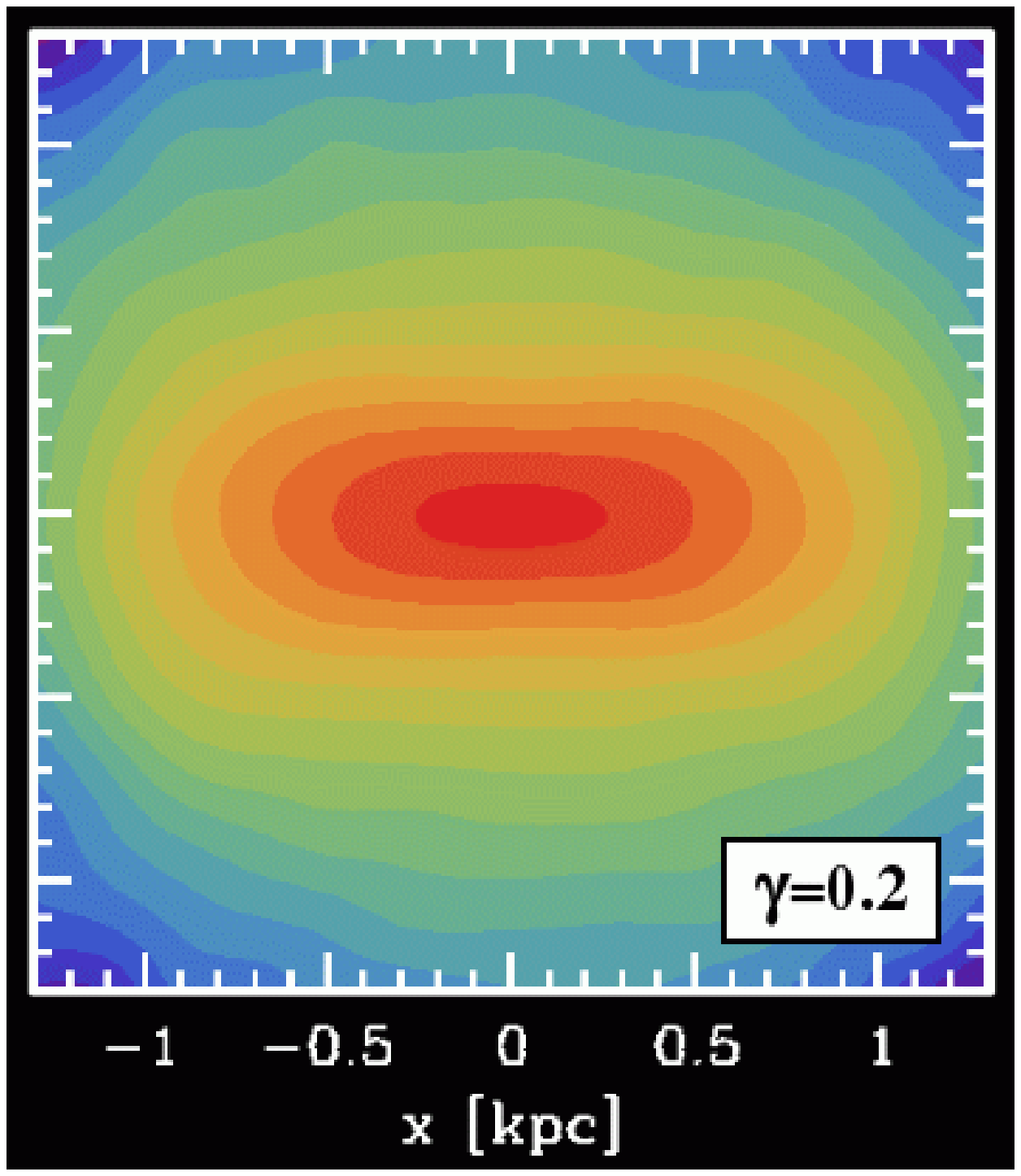}
\vspace{0cm}
\end{tabular}
\end{center}
\vspace{-0.3cm}
\caption{Surface number density distribution of stars in the
  simulated dwarfs.  Results are presented for orbits O1 (upper
  panels), O3 (middle panels), and O5 (lower panels). All stellar
  components are depicted at $t = 10$~Gyr, except for that in the
  upper rightmost panel (O1; $\gamma=0.2$) which is shown at $t =
  6.35$~Gyr (corresponding to the timescale of the last apocenter
  before the disruption of the dwarf galaxy). The asymptotic inner
  slope of the dwarf DM density profile, $\gamma$, is indicated in
  each panel. The images show the most non-spherical views along the
  intermediate axis of the stellar distribution, $y$ (where $x$ and
  $z$ denote the major and minor axes, respectively).  The stars are
  binned into $0.2~\rm{kpc} \times 0.2~\rm{kpc}$ fields perpendicular
  to the line-of-sight.  The contours correspond to the number of
  stars $N$ within each bin and are equally spaced by $0.2$ in $\log
  N$. The innermost contours are in the range $\log N = 3.6-4.6$
  depending on the case.
  \label{fig3}}
\end{figure*}


Our primary goal is to assess the likelihood and efficiency of
transformation into a dSph via tidal stirring. In accordance with
observational studies of LG dwarf galaxies \citep[e.g.,][]{Mateo98,
  McConnachie12}, we classify as {\it bona fide} dSphs only those
simulated systems whose final states are characterized by $V_{\rm
  rot}/\sigma_{\ast} \lesssim 0.5$ and $c/a \gtrsim 0.5$. We underline
that our theoretically derived values of $V_{\rm rot}/\sigma_{\ast}$
and $c/a$ are appropriate for meaningful comparisons with those of
observed dSphs (see \citealt{Kazantzidis_etal11a} for a thorough
discussion on this issue). In fact, the criterion $V_{\rm
  rot}/\sigma_{\ast} \lesssim 0.5$ for establishing dSph formation in
our experiments should be regarded as fairly conservative.  Indeed, we
measure $V_{\rm rot}$ around the minor axis of the stellar
distribution, which corresponds to observing the simulated dwarfs
perfectly edge-on.  Consequently, $V_{\rm rot}$ is {\it nearly}
equivalent to the maximum rotational velocity and, thus, the values of
$V_{\rm rot}/\sigma_{\ast}$ that we quote throughout the paper are
essentially the largest possible. Adopting a random line-of-sight
would result in smaller $V_{\rm rot}/\sigma_{\ast}$ values, indicating
even stronger and more complete transformations.

Columns 7 and 8 of Table~\ref{table:summary} list the final values of
$V_{\rm rot}/\sigma_{\ast}$ and $c/a$, respectively. These values
correspond to different timescales (from $\sim 4$ to $10$~Gyr) that
are determined by the presence of a well-defined, bound stellar
component of $\sim 0.7$~kpc in size. Column 9 reports whether a dSph
was produced according to the two criteria above. The notation
``dSph?''  in this column is introduced to account for observed dSphs
that exhibit $0.5 \lesssim V_{\rm rot}/\sigma_{\ast} \lesssim 1$ such
as Cetus \citep{Lewis_etal07}, Tucana \citep{Fraternali_etal09}, and
Andromeda II \citep{Ho_etal12}. Of the disky dwarfs initially embedded
in cuspy DM halos ($\gamma = 1$), only those on high-eccentricity
($r_{\rm apo}/r_{\rm peri} \gtrsim 5$) and small-pericenter ($r_{\rm
  peri} \lesssim 25$~kpc) orbits are transformed into objects with the
properties of dSphs \citep[see
also][]{Mayer_etal01,Kazantzidis_etal11a}. Most importantly, the
likelihood of dSph formation is enhanced significantly for shallow DM
density profiles ($\gamma<1$); in such cases, rotationally-supported
dwarfs on previously unfavorable low-eccentricity and/or
large-pericenter orbits are able to transform into dSph-like systems.
Figure~\ref{fig3} shows the surface density maps of the final stellar
components of the dwarfs and visually confirms these conclusions.

It is also worth emphasizing that, when $\gamma <1$, a dSph can be
created without significant mass loss in the stellar component (see
Figure~\ref{fig1}) because strong tidal shocks are not required to
induce transformation for $\gamma=0.6$ and $\gamma=0.2$. This finding
has interesting implications as it suggests that the presence or
absence of prominent tidal tails in present-day dSphs may not
constitute a robust observational constraint on the tidal stirring
model without prior knowledge of the inner DM density distributions of
such objects \citep{Mayer_etal02}.

Inspection of Columns 6 and 9 of Table~\ref{table:summary} illustrates
that dSph formation only occurs in conjunction with the development of
bar instabilities, highlighting a strong association between bar
formation and transformation into a dSph via tidal stirring \citep[see
also, e.g.,][]{Mayer_etal01,Klimentowski_etal09,Kazantzidis_etal11a}.
Lastly, Column 10 of Table~\ref{table:summary} lists the time elapsed
from the beginning of the simulation until dSph formation occurs (the
number of corresponding pericentric passages is included in
parentheses). The entries in this column indicate, that similar to the
likelihood, the efficiency of transformation into a dSph is
considerably increased for disky dwarfs embedded in DM halos with
shallow density profiles.

\section{Discussion}
\label{sec:discussion}

The present study is the first to elucidate the effect of the
asymptotic inner slope $\gamma$ of the dwarf DM density profile ($\rho
\propto r^{-\gamma}$ as $r \rightarrow 0$) on the tidal evolution of
dwarf galaxies consisting of baryonic components in the form of
stellar disks. We have shown that, regardless of orbit inside the
primary, rotationally-supported dwarfs embedded in DM halos with
core-like density distributions ($\gamma = 0.2$) and mild density
cusps ($\gamma = 0.6$) demonstrate a substantially enhanced likelihood
and efficiency of transformation into dSphs relative to their
counterparts with NFW-like steeper DM density profiles ($\gamma = 1$).
Such shallow DM distributions are akin to those inferred from the mass
modeling of observed LG dIrrs \citep[e.g.,][]{Weldrake_etal03} and
analogous galaxies outside of the LG \citep[e.g.,][]{Oh_etal11}. This
fact highlights tidal stirring as a plausible mechanism for the origin
of the morphology-density relation \citep[e.g.,][]{Mateo98}, an
essential constraint that any model for the LG must satisfy.

The following order-of-magnitude calculations offer insight into our
numerical results. In the impulse approximation, the energy injected
at each pericentric passage is given by $\Delta E \propto M_{\rm
  host}^2 M R^2 V_{\rm rel}^{-2}$, where $M_{\rm host}$ is the mass of
the host enclosed within $r_{\rm peri}$, $M$ denotes the mass of the
dwarf within a characteristic radius $R$, and $V_{\rm rel}$ is the
relative velocity of the two galaxies at the pericenter of the orbit
\citep[e.g.,][]{Binney_Tremaine08}. For a specific set of initial
orbital parameters, $M_{\rm host}$ and $V_{\rm rel}$ are fairly
similar for different $\gamma$. Thus, $\Delta E \propto M R^2$.
Moreover, by virtue of the virial theorem, the energy content of the
dwarf scales as $E \propto M^2/R$.  Hence, the fractional increase in
energy caused by the tidal shocks is $\Delta E / E \propto R^3 / M$.
At a given distance $R$ from the center of our initial disky dwarf
galaxies, decreasing cusp slopes correspond to smaller $M$, and thus
to larger $\Delta E / E$. This explains why the $\gamma = 0.6$ and
$\gamma = 0.2$ rotationally-supported dwarfs experience stronger tidal
shocks and augmented mass loss relative to their $\gamma = 1$
counterparts, leading to their enhanced morphological transformation
into dSphs. Considering adiabatic corrections to the energy change
predicted by the impulse approximation
\citep[e.g.,][]{Gnedin_Ostriker99} would only reinforce this
conclusion. Indeed, owing to the fact that such corrections are
inversely proportional to a power of the stellar orbital frequency
$\omega$ and $\omega$ is higher at a given radius for more
concentrated mass distributions, decreasing cusp slopes would
correspond to even larger $\Delta E / E$ compared to those predicted
by the impulse approximation.

The results presented in this paper demonstrate that the orbital
parameters and the DM density distributions of the progenitor
rotationally-supported dwarfs can independently determine the final
properties of dSphs formed via tidal stirring.  Therefore, the fact
that Fornax and Draco, for example, have roughly similar masses at
present, as inferred from their stellar velocity dispersions
\citep[e.g.,][]{Kazantzidis_etal04b}, but differ by about two orders
of magnitude in luminosity, could be explained in two ways.  One
possibility is that the predecessors of these two dwarf galaxies
acquired very different DM density profiles, for reasons related to
their formation history and not to the environment. Alternatively,
Fornax and Draco may have originated from systems that had comparable
DM and baryonic mass distributions, but displayed dissimilar tidal
evolutions and evolved differently because they entered their host
galaxy on different orbits, perhaps due to a different infall epoch
onto the primary. Other effects not included here, such as stripping
due to ram pressure and external radiation fields at high redshift,
may also affect the disky dwarfs differently depending on their orbits
\citep{Mayer_etal06,Mayer_etal07} and internal mass distributions,
increasing further the scatter in the properties of the resulting
dSphs.

The increased mass loss and rate of disruption experienced by dwarfs
embedded in DM halos with shallow density profiles have important
implications for alleviating the ``missing satellites problem''
\citep{Moore_etal99,Klypin_etal99}. These findings also suggest that
the dwarf luminosity and mass functions must be shifted lower relative
to those of {\LCDM} simulations that do not take into account baryonic
effects (such as outflows triggered by supernovae explosions; see
\citealt{Governato_etal10}) which can give rise to flattening of DM
cusps. Consequently, the recent claims regarding the agreement between
theory and observations of the luminosity function and the radial
distribution of satellites within MW-sized host halos
\citep[e.g.,][]{Maccio_etal10} may need revision.

Our investigation establishes tidal stirring as an even more prevalent
transformation mechanism than previously considered.  Indeed, earlier
work adopting cuspy DM density profiles for the progenitor disky
dwarfs has indicated that only orbits that are characterized by high
eccentricities ($r_{\rm apo}/r_{\rm peri} \gtrsim 5$), typical of CDM
structure formation models \citep[e.g.,][]{Diemand_etal07}, and small
pericentric distances ($r_{\rm peri} \lesssim 25$~kpc) are capable of
transforming rotationally-supported dwarfs into dSphs
\citep[e.g.,][]{Mayer_etal01,Kazantzidis_etal11a}. For a
transformation to occur, the same studies have also concluded that
$T_{\rm orb}$ should be short enough to allow the disky dwarf galaxies
to complete at least two pericentric passages inside their hosts.
These two requirements are probably not met for a number of LG dSphs,
a fact that questioned the widespread applicability of the tidal
stirring model.

For example, the distant dSphs Leo I, Cetus, and Tucana (residing in
the outskirts of the LG at several hundred kpc from the MW and M31;
e.g., \citealt{Caputo_etal99,McConnachie_etal05,Saviane_etal96}) are
likely moving on very wide orbits associated with extremely long
$T_{\rm orb}$. Moreover, proper-motion measurements have indicated a
fairly large pericenter for Leo I ($r_{\rm peri} \gtrsim 60$~kpc;
\citealt{Sohn_etal12}) and a low-eccentricity orbit with a similarly
great pericenter for the dSph Fornax \citep[e.g.,][]{Piatek_etal07}.
Our results suggest that the origin of all these dSphs can be
accommodated within the framework of the tidal stirring model (for
alternative formation scenarios of the distant LG dSphs and Fornax,
see, e.g.,
\citealt{Kravtsov_etal04,Sales_etal07,Ludlow_etal09,Kazantzidis_etal11b}
and e.g., \citealt{Yozin_Bekki12}, respectively). Indeed,
rotationally-supported dwarf galaxies on low-eccentricity and/or
large-pericenter orbits may be efficiently transformed into dSph-type
objects, provided that their DM halos possess shallow density profiles
(see Table~\ref{table:summary}). In fact, assuming core-like DM
distributions ($\gamma =0.2$) and small $r_{\rm peri}$, even a single
pericentric passage can induce transformation into a dSph. We note
that according to recent studies, Leo I has passed through the
pericenter of its orbit once \citep{Sohn_etal12} and the same
conclusion may also apply to Cetus and Tucana \citep{Teyssier_etal12}.

The present work confirms the previously reported strong association
between the development of bar instabilities in the disks of the
progenitor late-type dwarfs and the formation of dSphs
\citep[e.g.,][]{Mayer_etal01,
  Klimentowski_etal09,Kazantzidis_etal11a}. Bar-like structures should
thus be common among the less evolved dwarf galaxies in the LG as the
bar stage constitutes one of the longest phases in the transformation
process.  Recently, the first steps toward validating this fundamental
prediction of tidal stirring have been taken with the detection of
bars in a number of MW satellites, including Ursa Minor, Sagittarius,
and Carina \citep{Lokas_etal12b}.  Nonetheless, with the notable
exception of the Large Magellanic Cloud, the number of irrefutable
bar-like distortions observed in LG dwarfs is still relatively low
(see, however, \citealt{Klimentowski_etal09} for a discussion
pertaining to the intrinsic difficulties in identifying bars in the
dwarf galaxies of the LG).

Our simulations also indicate that none of the $\gamma = 1$ disky
dwarfs are disrupted by the tidal field of the primary galaxy; of
these, only systems on high-eccentricity ($r_{\rm apo}/r_{\rm peri}
\gtrsim 5$) and small-pericenter ($r_{\rm peri} \lesssim 25$~kpc)
orbits are transformed into objects with the properties of dSphs.
Interestingly, when $\gamma <1$, the dSph-like systems that survive
are generally on orbits with lower eccentricities and/or larger
pericenters compared to those of typical CDM satellites
\citep[e.g.,][]{Diemand_etal07}.  This novel finding provides a
natural explanation for the rather peculiar orbits of several classic
LG dSphs such as Fornax, Leo I, Tucana, and Cetus.

The degree of transformation into a dSph depends on both the number
and the strength of tidal shocks, which are determined by $T_{\rm
  orb}$ and $r_{\rm peri}$, respectively
\citep[e.g.,][]{Mayer_etal01,Kazantzidis_etal11a}. Specifically, short
orbital times and small pericentric distances, corresponding to orbits
associated with a large number of strong tidal shocks, produce the
strongest and most complete transformations.  Assuming that the
distant dSphs in the LG are products of tidal stirring and that their
progenitors have no bias towards significantly shallower than average
inner DM density distributions, we predict that these dwarfs should
exhibit higher values of $V_{\rm rot}/\sigma_{\ast}$ relative to those
of dSphs located closer to their hosts. Although tentative evidence of
intrinsic rotation exists in Leo I \citep[e.g.,][]{Koch_etal07}, Cetus
\citep{Lewis_etal07}, and Tucana \citep{Fraternali_etal09}, future
conclusive measurements of kinematics in these dSphs will serve to
validate (or falsify) this prediction.

Lastly, our approach neglects the effects of gas dynamics, star
formation, and chemical evolution. Definitive conclusions regarding
the efficiency of the transformation process clearly require more
sophisticated theoretical modeling, capable of capturing the
turbulent, multi-phase structure of the interstellar medium in dwarf
galaxies.  Investigating the tidal stirring of realistic dIrr
galaxies, such as those formed self-consistently in the cosmological
hydrodynamical simulations of \citet{Governato_etal10}, with recipes
of radiative cooling, star formation, and supernovae feedback would
thus be particularly valuable in this direction.

\acknowledgments

We acknowledge stimulating discussions with Giuseppina Battaglia,
Jonathan Bird, Simone Callegari, Alan McConnachie, and Se-Heon Oh.
S.K. is supported by the Center for Cosmology and Astro-Particle
Physics at The Ohio State University. This work was partially
supported by the Polish National Science Centre under grant
NN203580940. The numerical simulations were performed on the ``Glenn''
cluster at the Ohio Supercomputer Center (http://www.osc.edu).

\bibliography{ms} 

\end{document}